\documentclass[twocolumn]{aastex631}

\usepackage{amsmath}

\begin{document}

\title{Exploring neutral hydrogen in the radio MOlecular Hydrogen Emission Galaxies (MOHEGs) and prospects with the SKA}

\author[0009-0006-5549-5196]{Sai Wagh}
\affiliation{Indian Institute of Science, CV Raman Road, Bangalore-560012, India}

\correspondingauthor{Sai Wagh}
\email{saiwagh@iisc.ac.in}

\author[0000-0001-5829-1099]{Mamta Pandey-Pommier}
\affiliation{ CNRS/Laboratoire Univers et Particules de Montpellier, Universit´e de Montpellier \\
LUPM CC 072 - Place Eug`ene Bataillon 34095 Montpellier Cedex 5, France}
\affiliation{University Catholic of Lyon, 10, place des Archives 69288 Lyon Cedex 02, France}


\author[0000-0001-9829-7727]{Nirupam Roy}
\affiliation{Indian Institute of Science, CV Raman Road, Bangalore-560012, India}

\author[0009-0007-0120-5728]{Md Rashid}
\affiliation{Indian Institute of Science, CV Raman Road, Bangalore-560012, India}

\author[0000-0002-3971-0910]{Alexandre Marcowith}
\affiliation{ CNRS/Laboratoire Univers et Particules de Montpellier, Universit´e de Montpellier \\
LUPM CC 072 - Place Eug`ene Bataillon 34095 Montpellier Cedex 5, France}

\author{Chinnathambi Muthumariappan}
\affiliation{Indian Institute of Astrophysics, II Block, Koramangala, Bangalore 560034, India}

\author{Ramya Sethuram}
\affiliation{Indian Institute of Astrophysics, II Block, Koramangala, Bangalore 560034, India}

\author{Subhashis Roy}
\affiliation{National Centre for Radio Astrophysics - Tata Institute of Fundamental Research, \\
Ganeshkhind, Pune 411007, Maharashtra, India}

\author{Bruno Guiderdoni}
\affiliation{Univ Lyon, Univ Lyon1, Ens de Lyon, CNRS, Centre de Recherche Atsrophysique de \\
Lyon UMR5574, F-69230, Saint-Genis Laval, France}



\begin{abstract}
The empirical studies of cold gas content serve as an essential aspect in comprehending the star formation activities and evolution in galaxies. However, it is not straightforward to understand these processes because they depend on various physical properties of the Interstellar Medium. Massive FRI/II type radio galaxies rich in molecular hydrogen with less star formation activities are known as radio Molecular Hydrogen Emission Galaxies (MOHEGs). 
We present a study of neutral hydrogen gas-associated radio MOHEGs at redshifts $<$0.2 probed via the H~{\sc i} 21-cm absorption line. Neutral hydrogen is detected in  70$\%$ of these galaxies, which are located at a  distance of 8 - 120 kiloparsec from the neighboring galaxies. These galaxies show a scarcity of H~{\sc i} gas as compared to merging galaxies at similar redshifts. We found no strong correlation between N(H~{\sc i}), N$_\mathrm{H}$, and galaxy properties, independent of whether the H~{\sc i} is assumed to be cold or warm, indicating that the atomic gas is probably playing no important role in star formation. The relationship between total hydrogen gas surface density and star formation surface density deviates from the standard Kennicutt-Schmidt law. Our study highlights the importance of H~{\sc i} studies and offers insights into the role of atomic and molecular hydrogen gas in explaining the properties of these galaxies. In the upcoming H~{\sc i} 21-cm absorption surveys with next-generation radio telescopes such as the Square Kilometre Array (SKA) and pathfinder instruments, it may be possible to provide better constraints to such correlations.
\end{abstract}

\keywords{galaxies: absorption lines - galaxies: evolution - galaxies: star formation - galaxies: interactions}


\section{Introduction} \label{sec:intro}
Gas interactions between galaxies and the Intracluster Medium (ICM) play a crucial role in star formation activities and their evolution. Galaxies experience several interactions that influence their evolution over cosmic time, such as gravitational interaction and merger events among the nearest neighbors, tidal forces between neighboring galaxies,  ram pressure stripping, and strangulation in dense environments \citep{1974Toomre,1975Alladin,2012We,2008Kawata,2015Peng,2022Boselli}. These interactions can either trigger bursts of star formation and fuel the central supermassive black hole, stripping away their gas and leading to gradual loss of fuel and quenching of star formation. It eventually results in morphological changes from the stage of gas-rich star-forming galaxies with extended disks to quiescent ellipticals \citep{SFR2005, SFR2015,2013Vollmer,2019Pearson}. The lack of cold gas and quenching of star formation is often associated with heating the gaseous ICM to millions of Kelvin through feedback processes. For bright central galaxies(BCGs) in galaxy clusters, while a part of the molecular gas undergoes star formation, the rest of the gas is expected to accrete onto the central Super Massive Black Hole (SMBH), which ignites the Active Galactic Nuclei (AGN) radio-jet feedback activities. Feedback via AGN hosting SBMH at the centers of galaxies that accrete gas and release tremendous amounts of kinetic energy via radio-jets dissipates through shocks, resulting in turbulent heating of the molecular gas and reducing the star formation efficiency in the surrounding ICM \citep{2008Antonuccio,2018Cielo}. The mechanism of sweeping out gas from their host galaxies, i.e., negative feedback, prevents the cooling and condensation of gas onto galaxies, thereby regulating the amount of cold molecular gas available for star formation and the properties of the intergalactic medium \citep{Fabian1994, Fujita2022}. On the other hand, jet activities can also compress gas within the Interstellar Medium (ISM) and trigger star formation in dense cluster regions. This gives rise to a positive feedback mechanism, thereby enhancing star formation \citep{Silk2013, Maiolino2017}. The other mechanism of heating the ISM is feedback via explosive supernova events, which are responsible for driving powerful winds and shockwaves to expel cold gas from galaxies. However, supernova feedback is more relevant in lower-mass galaxies and can significantly impact their evolution \citep{2015Martizzi, 201Li}. In addition to this, the high-velocity gas outflows in mergers have been reported as a result of QSO(Quasi-Stellar Object) mode feedback, and it can majorly contribute to powering the AGN winds in mergers \citep{2008Narayanan,2011Rupke,2017Veilleux}.  The connection between these feedback processes responsible for heating the ICM and the availability of cold molecular gas reservoirs is crucial for developing a comprehensive picture of the influence of different environments on star formation and galaxy evolution \citep{Man2018}.
\vspace{0.03cm}
\par One of the substantial factors that governs star formation in galaxies is the availability of cold (atomic + molecular) gas.  
The molecular gas (H$_2$) distribution within the ISM directly impacts star formation, which further depends upon the reserve of atomic hydrogen (H~{\sc i}) available. In the ISM, H~{\sc i} leads to the formation of H$_2$ through an exothermic process. Theoretical models like Langmuir-Hinshelword(LH) and Eley-Rideal(ER) have been proposed to understand the process of H~{\sc i} to H$_2$ transformation. Interstellar dust grains and Polycyclic Aromatic Hydrocarbons (PAHs) act as catalysts in these processes; they act as surfaces over which hydrogen atoms can be readily absorbed and diffused, enabling the three-body reaction, which is much more efficient than the gas-phase reaction \citep{2013Vidali,2018Foley,2023Park,2020A&A...634A..42T}. In late-type galaxies, the Kennicutt-Schmidt (KS) law suggests a direct proportionality between the surface density of the star formation rate and the positive power of cold gas surface densities within their ISM \citep{KS1959,1963Schimdt, Ken1998}. Moreover, the atomic phase of hydrogen in galaxies acts as a crucial intermediary constituent in the baryon cycle, directly influencing structural evolution and star formation activities in these systems. Gas-rich galaxy disks hold extended atomic gas in their disk regions, susceptible to distortions during tidal interactions. The effects of such activities can be studied through observations of the H~{\sc i} 21cm transition \citep{1994Yun,2005Noordermeer,2009Chung, 2017Brown,2022Dutta}. Additionally, observations of atomic gas outflows in radio galaxies (RG) suggest that AGN radio jet feedback has a galaxy-scale impact on the host ISM\citep{Morganti2003, Morganti2005,2012ApJ...747...95G, 2021Schulz}. This implies that atomic gas is also influenced by chemical, radiative, and mechanical feedback associated with star formation activities\citep{1974Cox,1995Wolfire,2017Naab}.

\par The atomic ISM traced by the H~{\sc i}  21cm in galaxies exists in two different phases: the Cold Neutral Medium(CNM), having densities $\geq 10$ cm$^{-3}$ and spin temperatures $\leq 200$K, and the Warm Neutral Medium(WNM) with densities around $\equiv$ 0.1 - 1 cm$^{-3}$ and spin temperatures $\geq 5000$K \citep{1965Clark, 2003Wolfire,2014Saury}. The gas densities decide the physical conditions of the atomic phases in the ISM. Below N(H~{\sc i}) $ \leq$ 10$^{17}$ cm$^{-2}$  column densities, the gas is optically thin and is ionized by UV photons, giving rise to Lyman-$\alpha$ emission lines; the spectra are known to be the Lyman-$\alpha$ forest \citep{1998Rauch}. At the range of 10$^{17}$ cm$^{-2}$ $\leq$ N(H~{\sc i}) $\leq$ 10$^{20}$ cm$^{-2}$, gas becomes optically thick, and the core of Lyman-$\alpha$ is saturated. Such systems are called Lyman limit systems \citep{1991Bergeon}. At the same time, there are Damped Lyman-$\alpha$ absorbers (DLAs) having N(H~{\sc i}) $\geq 10^{20}$ cm$^{-2}$ where the Lyman-$\alpha$ line gets optically thick in its naturally broadened wings \citep{2005Wolfe}. A threshold neutral hydrogen column density of N(H~{\sc i}) $\sim$ 5 x 10$^{20}$ cm$^{-2}$ is required to begin with the formation of H$_2$, which helps to shield the gas against radiation from UV photons \citep{1967Stecher,1979Federman,1971Hollenbach,2011Kanekar}.
 Unlike H~{\sc i} emission lines, where the flux attenuates with distance, H~{\sc i} absorption features are independent of the distance of the source; rather, they primarily depend on the strength of the background radio continuum, the covering fraction ($f\mathrm{_c}$) of the source, and spin temperature T$\mathrm{_s}$ (K) \citep{2022Dutta}. Hence, the H~{\sc i} 21-cm absorption line studies can complement the emission line surveys to trace the evolution of the atomic gas component in and around galaxies. The  H~{\sc i} 21-cm optical depth-integrated over velocity ($\tau_{(\nu)} d\nu$) is proportional to the column density of neutral hydrogen, N(H~{\sc i}) cm$^{-2}$ given by,

 \begin{equation}
 \centering
      \mathrm{N_{HI}}   = 1.82 \times 10^{18} \cdot \frac{\mathrm{T_{s}}}{f\mathrm{_c}} \int{}^{} \tau_{(\nu)} d\nu \; \; (\mathrm{cm^{-2}})
 \end{equation}
 where T$\mathrm{_s}$ (K) is the spin temperature of the gas, $f\mathrm{_c}$ is the fraction of the background radio source covered by the absorbing gas \citep{2000Rohlfs}.

\par \cite{2007ApJ...668..699O} found enhanced molecular emission from pure rotational line emission L(H$_2$) $=$ (8.0 $\pm$ 0.4) $\times$ 10$^{41}$ erg s$^{-1}$ from a radio galaxy system 
(3C 326N). Many more galaxies observed with $\it Spitzer$ showed this enhanced H$_2$ emission, leading to their classification as MOlecular Hydrogen Emission Galaxies (MOHEGs). \cite{2013Cluver} investigated 23 High Compact Galaxies (HCGs) with $\it Spitzer$, 14 of which are MOHEGs.  Studies from the $\it Spitzer$ Infrared (IR) telescope have revealed H$_2$ pure-rotational emission lines detected from their warm (100-1500 K) molecular gas up to the redshift of z $<$ 0.22 with L(H$_2$)/L(PAH) ratio (0.03-4 or greater), up to a factor of $\sim$300 this ratio is greater than that of normal star-forming galaxies. These galaxies with FR-I and FR-II type radio morphologies are named radio MOHEGs \citep{Ogle2010}. The measured pure rotational and ro-vibrational H$_2$ transitions measured in these galaxies indicated H$_2$ luminosities in the range of 10$^{40}$-10$^{42}$ erg s$^{-1}$, that act as the primary coolant for the warm molecular phase of the ISM, characterized by short cooling timescale ($\sim$ 10$^4$ years) \citep{1999Le,2009Maret}. 
 Despite possessing a substantial amount of molecular hydrogen gas ($\sim$ 10$^{10}$ M$_\odot$), these galaxies exhibit inefficiency in star formation. This suggests that the excess molecular gas is kinematically unsettled and turbulent, hindering its cooling and star formation processes. 

There is uncertainty regarding the origin of large quantities of molecular gas in radio MOHEGs. Additionally, these galaxies are found either in dense galaxy cluster environments or in close interacting pairs or groups, expected to be stripped of their H~{\sc i} gas content. However, the observed high molecular mass in these galaxies indicates enhanced cooling flows driving gas into their central regions \citep{2007ApJ...668..699O}. Thus, to get a complete picture of the neutral and molecular hydrogen gas content in these galaxies as well as the impact of possible interactions on star formation, we carried out a study on a sample of 10 radio MOHEGs with H~{\sc i} absorption data available in the literature.
This paper presents a study of the distribution of neutral hydrogen gas in a sample of 10 radio MOHEGs, exploring the global H~{\sc i} absorption and total hydrogen gas properties. We also explore the correlation between hydrogen gas and various galaxy properties, presenting their relationships with star formation. The rest of this paper is organized as follows: Section $\S$ \ref{section2} describes the radio galaxy sample and provides details on H~{\sc i} absorption data. Section $\S$ \ref{section3} presents the results of the study of neutral hydrogen gas distribution in our radio MOHEGs sample, highlighting its association with galaxy properties and investigating the possibility of radio-jet-induced feedback heating the molecular gas. In Section $\S$ \ref{section4}, we discuss the potential role of upcoming telescopes such as SKA and its precursors in addressing the H~{\sc i} to H$_2$ phase related to MOHEG-type galaxies through deeper H~{\sc i} observations. Finally, $\S$ \ref{section5} summarizes our results and conclusions.
\newline Throughout this work, we consider a flat $\Lambda$-CDM cosmology with Hubble constant $H_o$ = 70 km s$^{-1}$ Mpc$^{-1}$, matter density parameter $\Omega_M$ = 0.30.

\setlength{\arrayrulewidth}{0.1mm}
\setlength{\tabcolsep}{1pt}        
\begin{table}
\caption{Observation details of the radio MOHEG catalog}
\centering
        \begin{tabular}{ccccc}
        \hline
        Source &  Type & z   & Observations  &  References \\
    
         (1) &(2) & (3) & (4) & (5) \\
         \hline
       Cen A	& I, TJ& 0.00183   & HIPASS &  \citet{2014Allison}\\
       3C 31	&  I,TJ  & 0.01701   &  VLA-C & \citet{2017Moss} \\ 
       3C 84	&  I, FD& 0.01756 & JLVA-A &  \citet{2023Morganti}\\ 
       3C 218&  I, TJ&0.05488   & VLA-A &  \citet{1995ApJ...442L...1D}
       \\
       3C 270 & I, TJ &0.00747   & VLA-C &  \citet{1994Jaffe}
       \\
3C 293	&  I, CSC& 0.04503   & WSRT &  \citet{2015GER}
\\
     3C 317	& I, FD & 0.03446  &  VLA-A &  \citet{1989AJ.....97..708V}
     \\
       3C 338	&  Ip, FD& 0.03035  &  GMRT &  \citet{2013MNRAS.429.2380C}
       \\
 3C 405	&II  & 0.05608  &  VLBA + &  \citet{2010Struve}
 \\    
& &  &  phased VLA & 
\\
       3C 433	&  II, X& 0.10160 &  Arecibo & \citet{2016Curran} 
       \\  

    \hline 
    \end{tabular}

     
         \small Notes -  Column 1: object name (NED). Column 2: radio morphologies according to \citealt{1974Fanaroff} classification; where I:FR-I type \& II: FR-II type, p: peculiar, TJ: twin jet, FD: fat double, CSC: compact symmetric core, DD: double-double, X: “X” shaped. Column 3:  redshift (NED). Column 4: H~{\sc i} observations. Column 5: references for H~{\sc i} observations 
    \label{table1}
    \end{table}

\setlength{\arrayrulewidth}{0.5mm}
\setlength{\tabcolsep}{8pt}        
\begin{table*}
\caption{Hydrogen gas in radio MOHEGs sample}

\centering
  \begin{tabular}{cccccccccccc} 
 \hline
        Source & N(H~{\sc i}) & $\mathrm{N_{H_2}}$ &$\mathrm{N_H}$&  && $\textit{f}_{\mathrm{{H_2}}}$&& F$_{178}$ &log $\nu $L$_{\nu178}$ & Nearest neighbor   \\
        
         &  (10$^{20}$)& (10$^{20})$ &(10$^{22})$& &  && &  &&distances ($\rho$)  \\
         & cm$^{-2}$ & cm$^{-2}$ &cm$^{-2}$  &I&II&III&IV&Jy& erg s $^{-1}$ & kpc\\
         (1) &(2) & (3) &(4) & (5)  &(6) & (7) &(8) & (9)  &(10)  & (11) \\
         \hline
       Cen A&5.70&1150&23.10&0.99&0.99&0.96&0.83&  1005 &40.12&8.25\\
       3C 31&  $<$0.60 & 88.3& $<$1.77&$<$0.99&$<$0.99&$<$0.99&$<$0.97& 18.3 &40.33&5.81\\
       3C 84	& 2.40 & $<$302&$<$6.06&$<$0.99&$<$0.99&$<$0.98&$<$0.93&   66.8&40.92&29.51\\
       3C 218&  7.00  &82.0&1.71&0.96&0.96&0.89&0.66&  225.7 &42.46&9.86\\ 
       3C 270 & 7.20 &$<$50.3&$<$1.08&$<$0.93&$<$0.93&$<$0.74&$<$0.49& 53.4   &40.07& 16.84 \\
       3C 293	&   14.50&215&4.41&0.97&0.97&0.86&0.72&   13.8   &41.07&119.85\\
       3C 317	 & $<$1.62&$<$20.3&$<$0.42&$<$0.96&$<$0.95&$<$0.85&$<$0.63&   53.4   &41.42&5.77\\
       3C 338	&  $<$5.59&$<$30.1&$<$0.65&$<$0.92&$<$0.92&$<$0.67&$<$0.42&  51.1   & 41.28&18.97\\ 
       3C 405	&   23.1&$<$200&$<$4.23&$<$0.95&$<$0.92&$<$0.75&$<$0.54&  9483  &44.10&10.72\\
       3C 433	&  2.30&310&6.22&0.99&0.99&0.99&0.98&   61.3   &42.46&29.62\\
         
    \hline  
    \end{tabular}
         
         \small Notes- Column 1: object name (NED). Column 2: average N(H~{\sc i}) of detections and 3$\sigma$ upper limits on non-detections for T$_\mathrm{s}$ of 100 K. Column 3: total $\mathrm{N_{H_2}}$ of detections and 3$\sigma$ upper limits on non-detections \citep{Ogle2010}.  Column 4: total hydrogen column densities. Column 5: molecular fraction computed for N(H~{\sc i}) assuming a spin temperature of  100 K, Column 6: 100 K to 160 K. Column 7: 250 K to 680 K. Column 8: 1100 K to 1500 K.  Column 9: 178 MHz flux densities \citep{Baars1977}. Column 10: radio luminosity at 178 MHz \citep{Ogle2010}. Column 11: nearest neighbor angular separation taken from NASA NED, we use the scale (NED WRIGHT Cosmological Calculator, \citep{2006Wright}) to convert the distances in kpc.
        

    \label{table2}
    \end{table*}

\begin{figure*}
\centering

	\includegraphics[width=1.0\columnwidth]{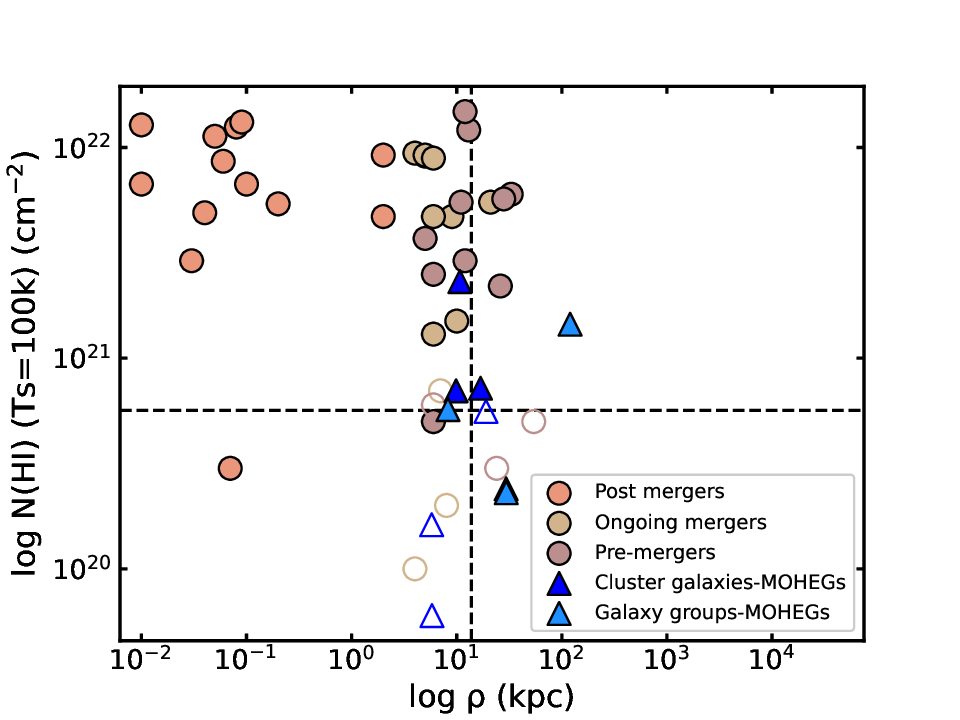}
 	\includegraphics[width=1.0\columnwidth]{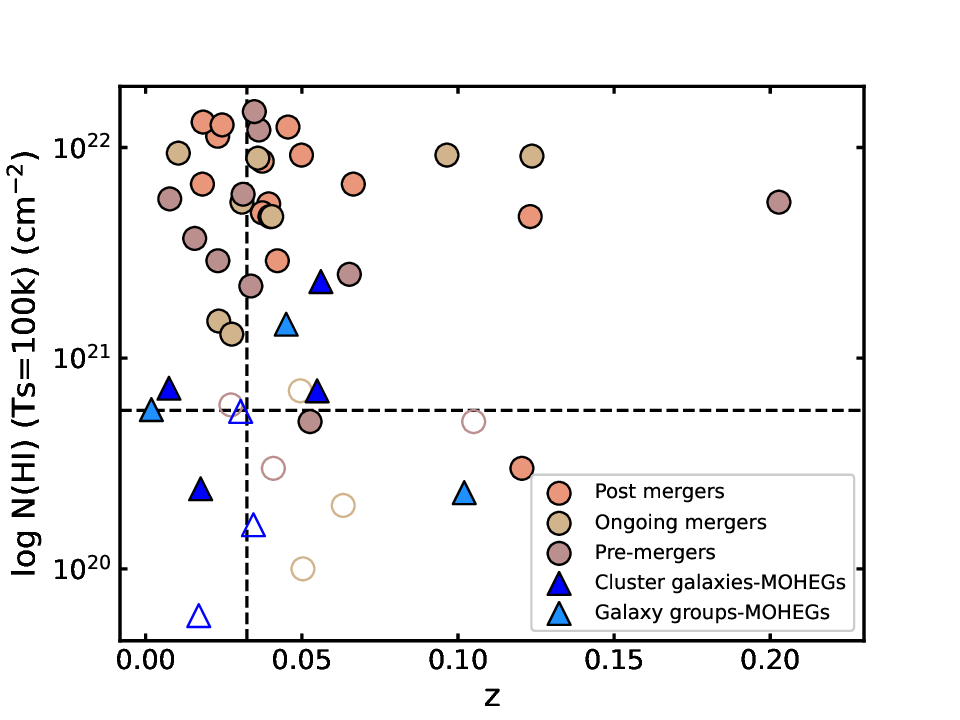}

    \caption{Comparison of neutral hydrogen column densities with the nearest neighbor distance (shown in the left panel) and redshift (shown in the right panel) between the radio MOHEG sample (represented as blue triangles) and a set of merger samples (represented as circles) sourced from \citep{2019MNRASDutta}. 
    The markers filled with color represent instances of H~{\sc i} detections in the dataset, while the unfilled markers indicate H~{\sc i} non-detections, considering 3$\sigma$ sensitivity levels. The dashed lines correspond to the median values extracted from the radio MOHEG sample, with medians of N(H~{\sc i}) $\leq$ 5.65 x 10$^{20}$ cm$^{-2}$, z = 0.03245, and $\rho$ = 13.78 kpc.}
    \label{fig:statNHI}

\end{figure*}    

\section{Sample and data selection } \label{section2}
 This study is based on a sample of radio MOHEGs with excess H$_2$ gas detected in $\it Spitzer$ IR Spectrograph observations \citep{Ogle2010}. The radio counterpart of these galaxies shows both FR-I and FR-II type morphology lying at z $<$ 0.2 (belonging to 3C RR and 3C R catalogs). Out of 17 H$_2$ detected MOHEG sources from \citep{Ogle2010}, we recovered H~{\sc i} absorption data for 10 sources from the existing literature and  GMRT, VLA, WSRT, and Arecibo archives as listed in Table \ref{table1}. The remaining seven sources, 3C 310, 3C 315, 3C 272.1, 3C 386, 3C 436, 3C 424, and 3C 326N, have not been observed earlier in the L-band for H~{\sc i} studies, and future observations for these sources will be proposed with existing and upcoming radio telescopes (e.g. SKA, VLA, MeerKat and uGMRT) to understand the atomic gas kinematics of radio MOHEGs. In this paper, we study the H~{\sc i} gas properties of 10 H$_2$ detected radio MOHEGs.
 The hydrogen gas (both H~{\sc i} and H$_2$) properties for our sample are listed in Table \ref{table2}. The H~{\sc i} column densities calculated are the average of column densities along adjacent lines of sight against multiple components of the continuum, however, for ARECIBO observations, the sources are not resolved due to the coarse resolution of 3.5 arcminutes offered by this instrument \citep{1990ApJ...351..503C,1978Knapp}. The total hydrogen column densities and molecular hydrogen fraction for our sample were computed using the following equations \citep{2017Winkel},

 \begin{equation}
 \label{eq:2}
 \centering
      \mathrm{N_H} = \mathrm{N_{HI} + 2\mathrm{N_{H_2}}} 
 \end{equation}
 and,
 \begin{equation}
 \label{eq:3}
 \centering
      \textit{f}_{\mathrm{{H_2}}} = \mathrm{\frac{2\mathrm{N_{H_2}}}{2\mathrm{N_{H_2}} + N_{HI}}}.  
 \end{equation}
The molecular hydrogen fraction ${f}_{\mathrm{{H_2}}}$ was calculated for these galaxies across a spectrum of spin temperatures ranging from 100 K to 1500 K. We aim to understand the change in molecular fraction with the varying spin temperature of the H~{\sc i} gas. To understand the interaction of these galaxies with surrounding environments in our sample, one way is to consider the distances from the cluster's center or the distances to the nearest neighbors. All of our sources have neighbors at a few kpc distances. Also, 7/10 sources reside inside clusters, and in that case, the distance from the cluster center may be a better proxy for studying the interaction effects. But expect two sources; all of the sources residing in the cluster have neighboring galaxies nearer than the cluster's center.
 Therefore, we choose the nearest neighbor distances for studying interaction effects with uniformity. We searched for the nearest neighboring galaxies at a similar redshift as radio MOHEGs in the NASA NED archive database. The maximum velocity difference between our sources and the nearest neighbors is 1388 km s$^{-2}$. We converted the projected angular separations to projected linear separations (in kpc) using the conversion scales obtained from Ned's Cosmological Calculator \citep{2006Wright}, as listed in Table \ref{table2}. The radio luminosities at 178 MHz are an excellent proxy for AGN radio-jet powers, as in this band, the total flux measured is synchrotron in origin dominated by emission from the jet region. Finally, the Star Formation Rates (SFRs) were computed from either the 7.7$\mu$m PAH, the 11.3$\mu$m PAH emission lines observed in the IR $\it Spitzer$ spectroscopic data, or the 70$\mu$m $\it Hershcel$ PACS(Photoconductor Array Camera and Spectrometer)/$\it Spitzer$ MIPS(Multiband Imaging Photometer) photometric data.
The median properties for our sample are: z = 0.0324, projected separation $\rho$ = 13.78 kpc, radio luminosity $\nu$L$_{\nu178}$ = 1.55 $\times$ $10^{41}$ erg s$^{-1}$,  and SFR = 0.52 M$_\odot$ yr$^{-1}$. We have also investigated the association between the gas column densities and global properties of the radio MOHEGs sample, as listed in Table \ref{table3}.

\section{RESULTS AND DISCUSSION} \label{section3}
The  H~{\sc i} 21-cm absorption lines are powerful tracers of cold gas in the central regions (at pc-scales), close to the central black hole of galaxies and up to tens of kpc tracing interaction and mergers into galaxy groups as well as the intervening neutral gas clouds in foreground galaxies, tails, and filaments \citep{Morganti2018,2022Dutta}. H~{\sc i} absorption in radio sources is mostly detected against the central regions, allowing us to probe the gas properties in the sources' inner regions and the AGN structures. Below, we discuss the results of our investigation of the H~{\sc i} absorption properties of MOHEGs galaxies, the influence of the environment on their merger activities, and the impact of radio mode feedback on the gas properties.

\subsection{Distribution and kinematics of H~{\sc i} gas}


\begin{figure*}
    \centering
    \includegraphics[width=2.2\columnwidth]{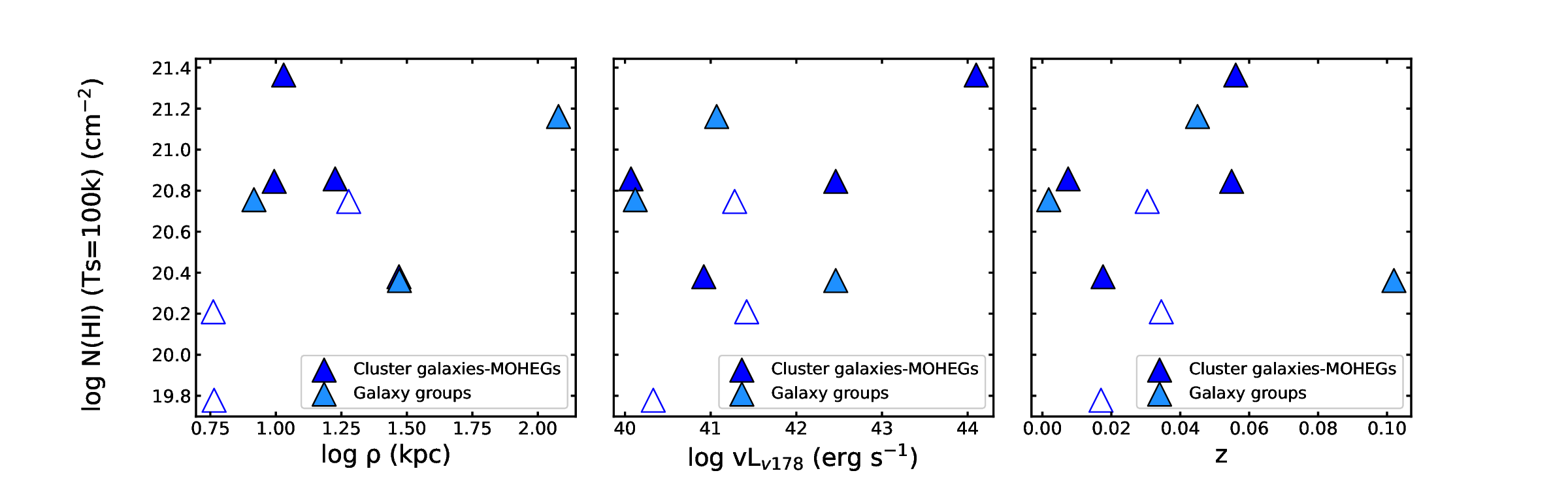}
    \caption{The inferred N(H~{\sc i}) values for radio MOHEGs are presented with projected separation, 178 MHz radio luminosity, and redshift, sequentially from left to right. Filled symbols represent cases of detection, while unfilled symbols denote 3$\sigma$ upper limits. Based on Kendall Tau's non-parametric test, N(H~{\sc i}) exhibits no correlation with these parameters, as detailed in Table \ref{table3}.}
    \label{fig:corrnhi}
\end{figure*}

 \setlength{\arrayrulewidth}{0.5mm}
\setlength{\tabcolsep}{5pt} 
\begin{table*}
\caption{Correlation analysis between the hydrogen gas and various galaxy properties of radio MOHEG sample}
  \centering
  \begin{tabular}{ccccccccccccccccccccccc}
  \hline
    Variables&&&& N(H~{\sc i}) &&&&& $\mathrm{N_H}$ &&&&&$\textit{f}_{\mathrm{{H_2}}}$ &&&&& SFR\\
    &&& $\tau_k$ &$p_k$& $m_t$    &&& $\tau_k$ &$p_k$&$m_t$ &&&$\tau_k$ &$p_k$& $m_t$ &&& $\tau_k$ &$p_k$& $m_t$ \\
     (1)&&&(2)&(3)&(4)&&&(5)&(6)&(7)&&&(8)&(9)&(10)&&& (11)&(12)&(13) \\
    \hline
    $\rho$&&&0.20&0.47&0.3274&&&0.15&0.56&0.7685&&&0.04&0.92&0.0036&&& 0.46&0.06&0.2786 \\
    $\nu$L$_{\nu178}$&&&0.08&0.78&0.0736&&&0.00&1.00&-0.0072&&&-0.11&0.69&-0.0041&&& 0.17&0.51&0.2255\\
    z&&&0.13&0.64&8.1013&&&0.11&0.69&0.4279&&&0.00&1.00&0.0174&&& 0.28&0.26&12.2154\\

    \hline
  \end{tabular}
   
         \small Notes- Column 1: Correlation between the properties ($\rho$, L$_{178}$ and z) of the radio MOHEGs sample and N(H~{\sc i}), $\mathrm{N_H}$, ${f}_{\mathrm{{H_2}}}$, and SFR tested using non-parametric correlation tests. Columns 2,5,8,11: Kendall-Tau rank correlation coefficient. Columns 3,6,9,12: the p-values. Columns 4,7,10,13: Akritas-Theil-Sen slope. These tests are done with R software using \textbf{cenken} function from \textbf{NADA} package, especially used for censored data points in survival analysis.
       
    \label{table3}
\end{table*}
The average column density for H~{\sc i} disks in spiral galaxies is approximately 6 $\times$ 10$^{20}$ cm$^{-2}$ \citep{2016Wang}. Meanwhile, early-type galaxies tend to have H~{\sc i} column densities measured maximum up to 10$^{20}$ cm$^{-2}$ \citep{Serra2012}. In a sample of nearby radio galaxies, these values range from 10$^{20}$ to 10$^{21}$ cm$^{-2}$ with an assumed spin temperature of T$\mathrm{_s}$= 100 K \citep{2010Emonts}.
Radio MOHEGs showcase a wide range of H~{\sc i} column densities spanning from approximately $\approx$ 6 × 10$^{19}$ cm$^{-2}$ to  2.31 × 10$^{21}$ cm$^{-2}$. Hence, the radio MOHEGs exhibit column densities, standing in contrast to spirals and early-type galaxies. Nevertheless, the recorded H~{\sc i} column densities in radio MOHEGs remain lower compared to those found in Seyfert galaxies, where column densities can reach up to  10$^{22}$ cm$^{-2}$ \citep{Morganti2018}.
Among the 10 galaxies in our sample, neutral hydrogen is detected in 7 galaxies, with H~{\sc i} column densities reaching as low as 10$^{20}$ cm$^{-2}$  {\it highest} temperature of H$_2$ estimated for that source \citep{Ogle2010}. Figure \ref{fig:statNHI} shows the column density of H~{\sc i} versus the nearest neighbor distance (left) and redshift (right) for radio MOHEGs and a sample of mergers \citep{2019MNRASDutta} for comparison. 

From the left panel of Figure \ref{fig:statNHI}, we see that radio MOHEGs have nearest neighbor distances of 5-120 kpc, which fall in the same range as the projected separation of pre-mergers and ongoing mergers. Hence, we suggest that radio MOHEGs are engaged in ongoing mergers and pre-merger stages. Moreover, the median H~{\sc i} column densities are 5.65 x 10$^{20}$ cm$^{-2}$, but upper detection limits suggest that the H~{\sc i} column densities can be much lower in radio MOHEGs. The median N(H~{\sc i}) value for the mergers sample lies at $\sim$ 5 $\times$ 10$^{21}$ cm$^{-2}$, which is approximately nine times higher than the median N(H~{\sc i}) value observed in radio MOHEGs. From the above results, we infer that radio MOHEGs exhibit a deficiency in atomic hydrogen compared to merging galaxies at similar redshifts. Figure \ref{fig:statNHI} right panel explains that there is no preferential distribution of N(H~{\sc i}) with redshift between radio MOHEGs and mergers sample. However, the redshift range is too small to rule out any such trend conclusively.
In addition, previous studies have shown that galaxies residing in galaxy clusters tend to have less neutral hydrogen content than the similar types of galaxies lying in the isolated fields; mechanisms like ram pressure stripping and tidal interactions could explain this H~{\sc i} deficiency \citep{2016Denes}.

\begin{figure*}
    \centering

        \includegraphics[width=2.2\columnwidth]{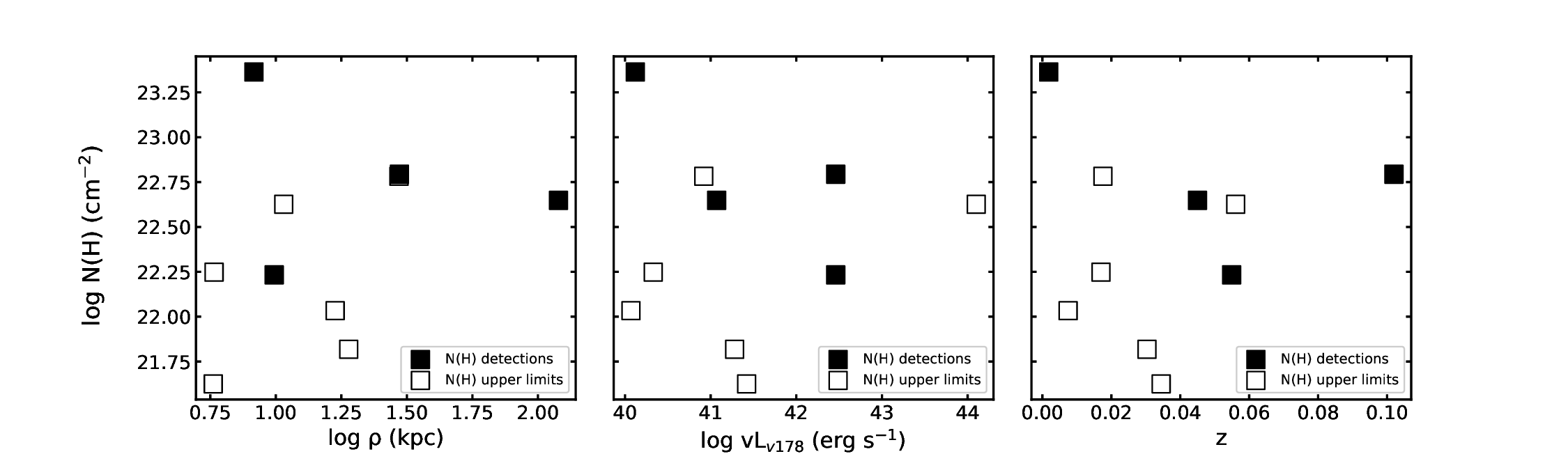}

    \caption{The inferred N$\mathrm{_H}$ values for radio MOHEGs are presented versus projected separation, 178 MHz radio luminosity, and redshift, sequentially from left to right. Filled symbols represent cases of detection, while unfilled symbols denote 3$\sigma$ upper limits. Based on Kendall Tau's non-parametric test, N$\mathrm{_H}$  exhibits no correlation with these parameters, as detailed in Table \ref{table3}. }
    \label{fig:corrth}
\end{figure*}

\subsection{ Hydrogen gas and galaxy properties}
We carried out non-parametric correlation tests to study the association between hydrogen gas properties and galaxy properties such as radio jet luminosities, redshift, and the effects of the host environment. The results of Kendall Tau's correlation coefficient ($\tau_k$) and the probability values (p-values, $p_k$) verify if the correlation is statistically significant. The association and probability values between the physical parameters are presented in Table \ref{table3}, along with Akritas-Theil-Sen (ATS) slopes ($m_t$), aims to find the best fit, giving an unbiased estimate of correlation trend magnitude with better asymptotic efficiency than the least-squares estimation. We do not find any significant correlation between the global H~{\sc i} or H~{\sc i} + H$_2$ gas profiles with those of the galaxy properties. We note that the results presented in Table~\ref{table3} are using N(H~{\sc i}) estimated assuming an average T$\mathrm{_s}$ = 100 K. However, qualitatively, the lack of correlations is found to be independent of that assumption. If we consider the extreme scenario that the absorptions instead arise from H~{\sc i} with T$\mathrm{_s}$ in the range of $1100 - 1500$ K (i.e., same as the highest temperature of H$_2$ for the corresponding source from \citep{Ogle2010}), there is no qualitative change in our results. It suggests that comprehending the triggering or quenching of nuclear star formation and its dependence on hydrogen gas in central regions of these galaxies based on H~{\sc i} absorption properties poses a challenge.  
The maximum correlation measured between all the derived quantities and the galaxy properties between the SFRs and the nearest neighbor distances is 0.46, which implies a weak correlation. In this case, we have obtained a p-value of 0.06, which is very close to 0.05, and we suggest that these two quantities are possibly weakly correlated. This would imply a weak dependence of the star formation activities on the surrounding environments.

As presented in Table \ref{table3}, we observe no correlation between the interaction distance on the H~{\sc i} column densities in our sample, albeit galaxy collisions or tidal interactions from neighboring galaxies that should make the ISM gas-rich. As seen in the mergers sample of \citep{2019MNRASDutta}, N(H~{\sc i}) has a weak dependence on the projected separations, although as seen in Figure \ref{fig:corrnhi} (Panel 1),  we see no such association here.
Further, we expect that the AGN radio jet feedback should expel the gas from the central regions of galaxies, making them deficient in neutral gas. We find no strong association of N(H~{\sc i}) with the radio jet luminosity evident from a high p-value of $\sim$ 0.8 in the radio MOHEGs sample, as seen in Figure \ref{fig:corrnhi} (Panel 2). However, \cite{2010Nes} and \cite{2012ApJ...747...95G} inferred that AGN feedback drives H~{\sc i} outflows in samples of molecular gas-rich radio galaxies. Ionized and atomic gas is well coupled dynamically and outflowing at comparable velocities in these systems. 3C 293, one of
the radio MOHEGs, shows broad blueshifted H{\sc i} absorption, indicating fast outflow of neutral gas that could originate near AGNs \citep{Morganti2003}. Also, ionized gas outflows from this system are reported \citep{2016Mahony}. However, we do not know how the molecular gas is affected \citep{2012ApJ...747...95G}. Until now, most observations of neutral hydrogen in galaxies have been conducted with limited sensitivity and narrow frequency ranges, constraining the detection of high-velocity H{\sc i} outflows. Detailed kinematic investigations of H{\sc i} can be conducted with high resolution, sensitivities, and wider bandwidths, enabling the identification of broader and shallower absorption features. Such AGN-induced outflows draw out the neutral gas from central regions in such systems, and the part of bulk kinetic energy outflowing neutral gas can be transferred to molecular gas heating through shocks in the ISM. No H{\sc i} or H$_2$ outflows have been reported for our sample so far (except for 3C 293); it will be intriguing to see if AGN-induced neutral gas outflows are also responsible for heating the molecular gas in radio MOHEGs. For two radio MOHEGs, 3C 293 and 3C 338, restarted radio jet activities have been reported \citep{ 2007Gentile,2013Mahony}, but none of the other sources have broadband measurements to study the interplay between jets and the environment. We also do not observe a correlation of total hydrogen gas column densities ($\mathrm{N_H}$) with the other properties of radio MOHEGs; refer to Figure \ref{fig:corrth}. However, understanding the total hydrogen content in such sources can provide valuable information about the ongoing and future star formation capacities. 
 The distinction between the scatter in the data points and the fit could arise due to the limited scope of our sample, where the extent of observations significantly affects the correlation analysis. Nevertheless, studying the effects of negative and positive AGN feedback in radio MOHEGs, which introduce turbulence within the ISM and potentially govern the H~{\sc i} content, is essential.

\subsection{Molecular fraction and SFRs}
To gain insights into the combined presence of H~{\sc i} and H$_2$ gas within the interstellar medium (ISM), we examined the molecular hydrogen fraction ($\textit{f}_{\mathrm{{H_2}}}$), within the radio MOHEG sample. This fraction reveals the proportions of H$_2$ and other molecular gases in the ISM. By investigating the dependence of radio MOHEG properties on this molecular fraction, we aim to enhance our understanding of the efficiency of converting H~{\sc i} into H$_2$ and its effects on star formation processes.
In the context of radio MOHEGs, when N(H~{\sc i}) is computed using T$_\mathrm{s}$ of 100 K, the average molecular fraction is $\leq$ 0.96. This fractional value implies that the hydrogen gas in radio MOHEGs is dominated by molecular gas. As listed in Table \ref{table2}, molecular fraction diminishes when neutral gas temperature varies with the temperature range where warm molecular gas is found. 
Furthermore, the correlation between molecular hydrogen fraction and SFRs within radio MOHEGs implies no association, characterized by a correlation coefficient of 0.22. A robust power-law relationship has been identified between the local molecular gas fraction and the pressure of the ISM among nearby galaxies, rich in both H~{\sc i} and H$_2$ gas, as well as dwarf galaxies \citep{2004Blitz,2006Blitz}. A study by \citet{2008Leroy} focused on 23 H~{\sc i} galaxies from THINGS, confirming a power-law connection between  $\textit{f}_{\mathrm{{H_2}}}$ and interstellar gas pressure, characterized by a power-law index of 0.8. However, our results signify no molecular fraction dependence on the galaxy properties of radio MOHEGs.

 \begin{figure*}
 \centering
    \includegraphics[width=2.2\columnwidth]{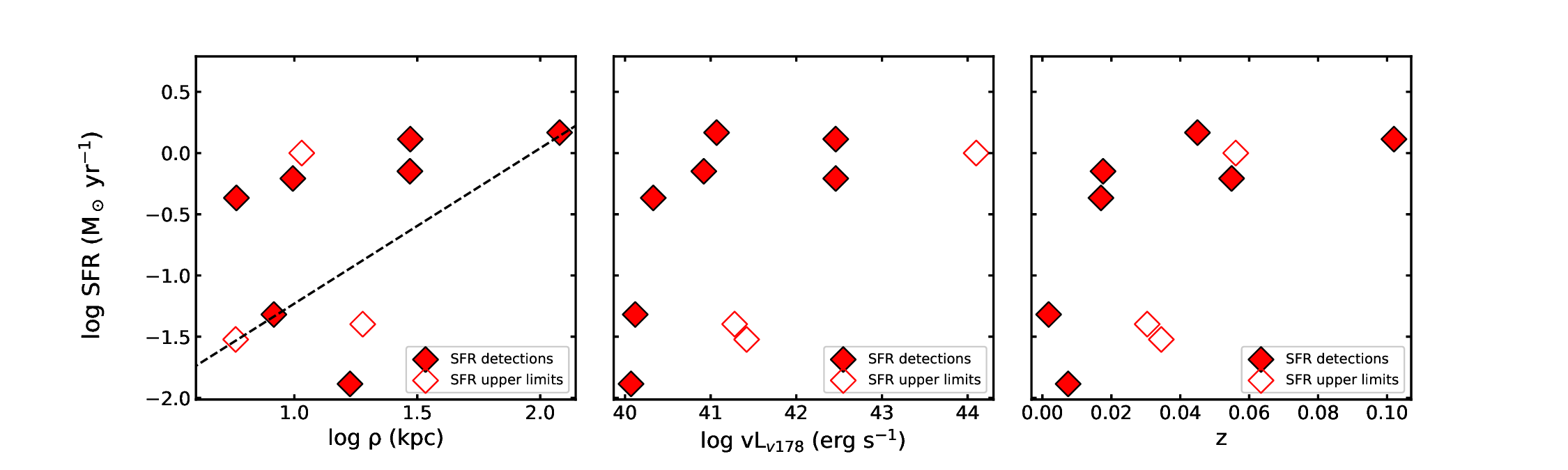}
\caption{Star formation rates in radio MOHEGs in correlation with nearest neighbor distances (Panel 1), 178 MHz luminosities (Panel 2), and redshift (Panel 3). The dashed line in Panel 1 represents the best line fit obtained from the Theil-Sen slope estimator for the possible correlation. Filled symbols represent cases of detection, while unfilled symbols denote upper limits. }
\label{fig:corrSFR}
\end{figure*}

In Figure \ref{fig:corrSFR}, we present the correlation between the SFRs and the physical properties of the radio MOHEGs sample. The SFRs exhibit a weak correlation with the nearest neighbor distances ($\tau_k$ = 0.46), suggesting that close-range galaxy interactions might lead to a decrease in star formation activities within radio MOHEGs due to the potential heating of molecular gas. This could be influenced by the negative feedback resulting from interactions and collisions with neighboring galaxies \citep{1998Kennicutt, 2018Pan}.
In Panel 2 of Figure \ref{fig:corrSFR}, despite a relatively low correlation value($\tau_k$ $\sim$ 0.17), the higher values of SFRs tend to align with higher values of radio luminosities. We note that galaxies with greater mass might exhibit elevated radio luminosities and increased star formation rates (SFR), but it's not necessarily indicative of luminosities directly influencing star formation activities.

\subsection{Star formation relations }
To comprehend the star formation activities in radio MOHEGs, we investigated the relations between the surface density of star formation ($\mathrm{\Sigma_{SFR}} $) and hydrogen gas surface densities.
 The empirical relationship that characterizes the correlation between the star formation density and hydrogen gas surface density in galaxies is known as the Kennicutt-Schmidt law (KS law) of star formation \citep{KS1959}, given by equation \ref{eq:5}.
 \begin{equation} \label{eq:5}
     \mathrm{\Sigma_{SFR}} = \mathrm{A} \Sigma_{\mathrm{gas}} ^{\mathrm{N}}
 \end{equation}
 where ($\mathrm{\Sigma_{SFR}} $) is star formation rate density in (M$_\odot$ yr$^{-1}$ kpc$^{-2}$), $\Sigma_{\mathrm{gas}}$ is the gas surface density $(\mathrm{M_\odot pc^{-2} })$, N is the power law index and 
A is a  proportionality constant.
 In normal star-forming galaxies, the KS law governs star formation by a power law scaling with a positive index (N$\sim $1.4), implying that the disk-averaged star formation rate per unit area has a linear relationship with the total surface gas densities (atomic and molecular gas) in logarithmic scale \citep{Ken1998}.
 We derived the SFRs (M$_\odot$ yr$^{-1}$) from the 7.7 $\mu$m, as well as the 11.3 $\mu$m PAH emission lines through the SFR relation given by \citep{Ogle2010} and converted it to $\mathrm{\Sigma_{SFR}} $ (M$_\odot$ yr$^{-1}$ kpc$^{-2}$) assuming the slit width area of 3.7'' $\times$ 10.0" which was used to extract the IR spectra at 10$\mu$m. We also used 70$\mu$m line luminosities from PACS \& MIPS photometric data \citep{2005Shi,2014MEL,2016Lanz} for calculating SFRs, using the relation derived by \citep{2010Calzetti} and converted it to $\mathrm{\Sigma_{SFR}} $ (M$_\odot$ yr$^{-1}$ kpc$^{-2}$) assuming circular aperture used for photometry of radii 12" for PACS and 30" for MIPS. We use equation \ref{CD2SD}, 
 \begin{equation} \label{CD2SD}
     \Sigma_{\mathrm{H_I + H_2}}  (\mathrm{M_\odot pc^{-2} })= \frac{\mathrm{N_{HI} + 2N_{H_2}}}{ 1.25 \times 10^{20} \mathrm{cm^{-2}}}
 \end{equation}%
 to convert column densities to surface densities. To maintain the consistency with \citep{Ken1998}, we parameterize gas surface densities ($\Sigma_{\mathrm{H_I + H_2}}$) in terms of H~{\sc i} surface densities.  

\begin{figure*}
    \centering
    \includegraphics[width=1\linewidth]{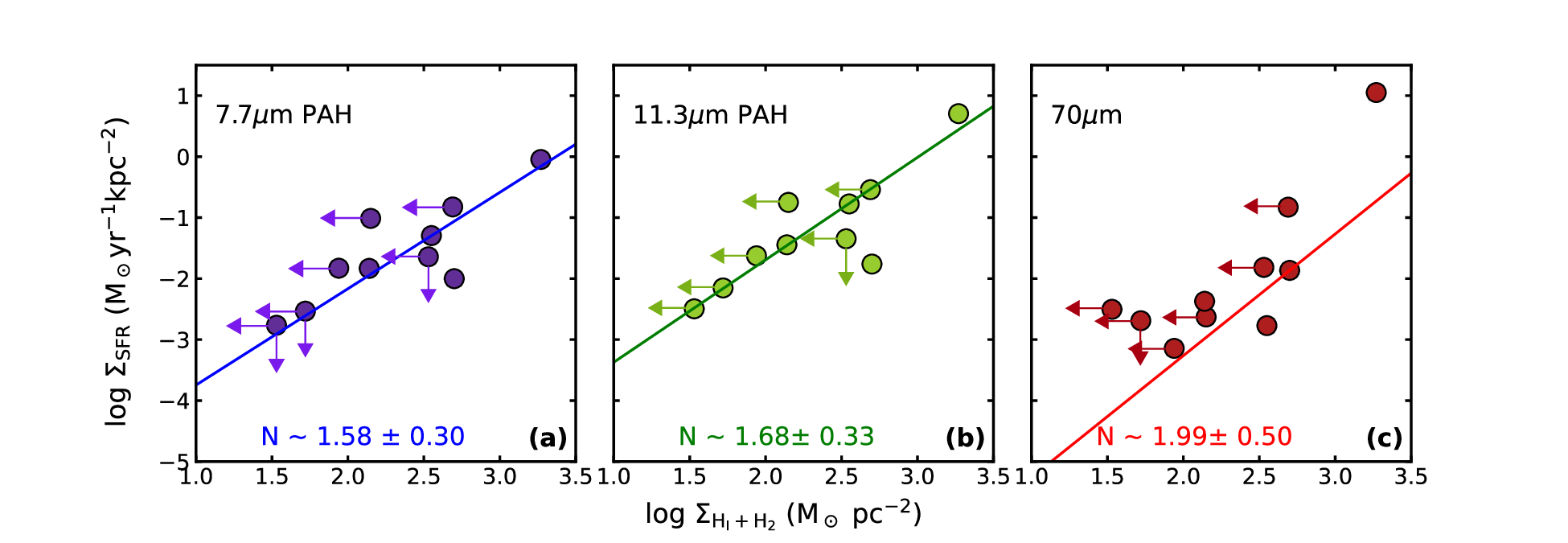}
    
    \caption{Comparison of KS relation for radio MOHEGs based on different tracers of SFRs. SFRs are computed using the luminosity of the (a) 7.7 $\mu$m PAH luminosities, (b) 11.3 $\mu$m PAH luminosities, and (c) 70 $\mu$m luminosities. N is the slope value for the best line fit estimated from the Theil-Sen slope estimator. Arrows represent upper limits on the x and y axes.}
    \label{fig:KSLAWcomparison}
\end{figure*}

From Figure \ref{fig:KSLAWcomparison}, we see that SFRs derived from all three tracers are comparable with slightly better correlation (less scatter) for the SFRs derived from the 7.7$\mu$m and the 11.3$\mu$m data for our sample. Also, the variation in slope is within the error bars, and no significant change in star formation densities is observed. \cite{Ogle2010} suggests that 11.3$\mu$m PAH-based SFRs should be used cautiously and considered upper limits. Therefore, we consider SFRs based on 7.7$\mu$m PAH lines to study the KS relation in our sample. 
In radio MOHEGs, there is a very weak correlation of the total hydrogen gas surface densities ($\Sigma_{\mathrm{H_I + H_2}}$) with the SFR densities ($\mathrm{\Sigma_{SFR}} $) calculated form 7.7$\mu$m PAH lines. The correlation coefficient value $\tau_k$ is 0.24 with a p-value of 0.32. The association between SFR densities and gas surface densities vanishes if we consider only the neutral gas surface densities with $\tau_k$ = 0.06, which agrees well with earlier studies on molecular-rich spiral galaxies that total cold gas (H~{\sc i} + H$_2$) densities or molecular gas densities influence star formation rates significantly rather than neutral gas \citep{2002Wong}. In addition, a tighter correlation is observed between SFR densities ($\mathrm{\Sigma_{SFR}} $) and molecular gas surface densities ($\Sigma_{\mathrm{H_2}}$) in radio MOHEGs with $\tau_k$ = 0.33 with a p-value of 0.17.

\begin{figure}
    \centering
    \includegraphics[width=1\columnwidth]{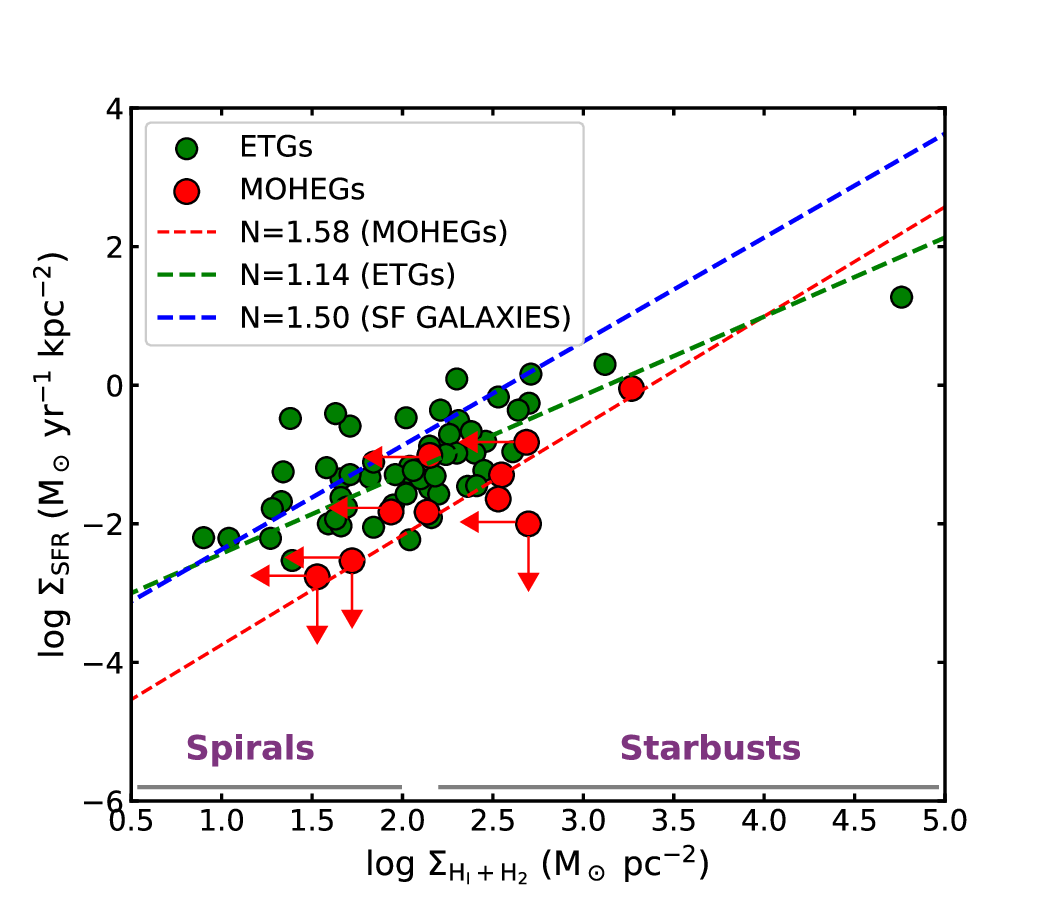}
    
    \caption{The Kennicutt-Schmidt law of star formation with the hydrogen gas (atomic + molecular) surface densities for normal star-forming and starburst galaxies. The power law fit for star-forming (SF) galaxies is taken from \citep{2021Kennicutt}, and ETGs from \citep{2014Davis} compared with the star formation relations in radio MOHEGs. The surface density of the gas is determined from C0 (2-1) emission for ETGs and SF galaxies, while for radio MOHEGs, it is determined from H$_2$ rotational line transitions \citep{Ogle2010}. Red arrows represent upper limits on the x and y axes. }
    \label{fig:KSLAW}
\end{figure}

In Figure \ref{fig:KSLAW}, we present a comparative analysis of radio MOHEGs alongside early-type galaxies (ETGs) to determine the relationship between these galaxy types and the KS law. We do not study star-forming relations with molecular gas because radio MOHEGs are H$_2$ dominated, and it would follow the same trend as with total gas densities. 
The spirals and starburst galaxies from the sample \citep{Ken1998} tend to lie on the empirically established KS law with an index of N = 1.41 $\pm$ 0.07 of the power law fit. However, \cite{2021Kennicutt} established a better power law fit for combined KS law for 119 galaxies, with a slope N = 1.50 $\pm$ 0.05, steeper than found by \citep{Ken1998}. 
 ETGs deviate from the combined KS relation, the best-fit line with a slope of 1.14 $\pm$ 0.04 for a sample of ATLAS$\mathrm{^{3D}}$ survey of ETGs. The ETGs, compared to spirals and starburst galaxies, have comparably lower average SFR surface densities \citep{2014Davis}. Radio MOHEGs exhibit a wide range of gas surface densities from 33-1844 M$_\odot$ pc$^{-2}$. In terms of the gas surface densities, radio MOHEGs reside between the low and high extremes of gas densities observed in starburst galaxies and normal spirals from Kennicutt's sample \citep{Ken1998}, respectively. Our analysis suggests that radio MOHEGs consistently deviate from the combined KS relation, indicating lower star formation surface densities than the central regions of normal star-forming galaxies and ETGs. This result emphasizes that radio MOHEGs emerge as a unique class of galaxies with an indication of lower star formation densities despite the abundance of molecular gas, as seen in Figure \ref{fig:KSLAW}. For radio MOHEGs, we report a star formation relation (SFR derived from 7.7$\mu$m) with total hydrogen surface densities with a power law index of N $\simeq$ 1.579 $\pm 0.305$ and A $\simeq$ -5.318 $\pm 0.726$ (computed using Akritas-Theil-Sen non-parametric line estimator using NADA package in R software)and can be written as,
\begin{align}
       \mathrm{log [\Sigma_{SFR}} \mathrm{(M_\odot yr^{-1} kpc^{-2})}] = (1.579 \pm 0.305) \times  \nonumber \\
\mathrm {log}[\Sigma_{\mathrm{H_I + H_2}}(\mathrm{M_\odot pc^{-2}})] - (5.318 \pm 0.726)
\end{align}
 We also infer that using different tracers for SFRs does not alter the position of radio MOHEGs in the KS law relation. Molecular gas mass derived from H$_2$ rotational lines and CO observations may have different systematics and may cause some offset in the KS law relation. It will be worth checking with a larger sample in the future if the offset seen here arises only due to such systematics.

\section{ H~{\sc i} absorption surveys and prospects with SKA } \label{section4}
\begin{figure*}
 \centering
    \includegraphics[width=2.2\columnwidth]{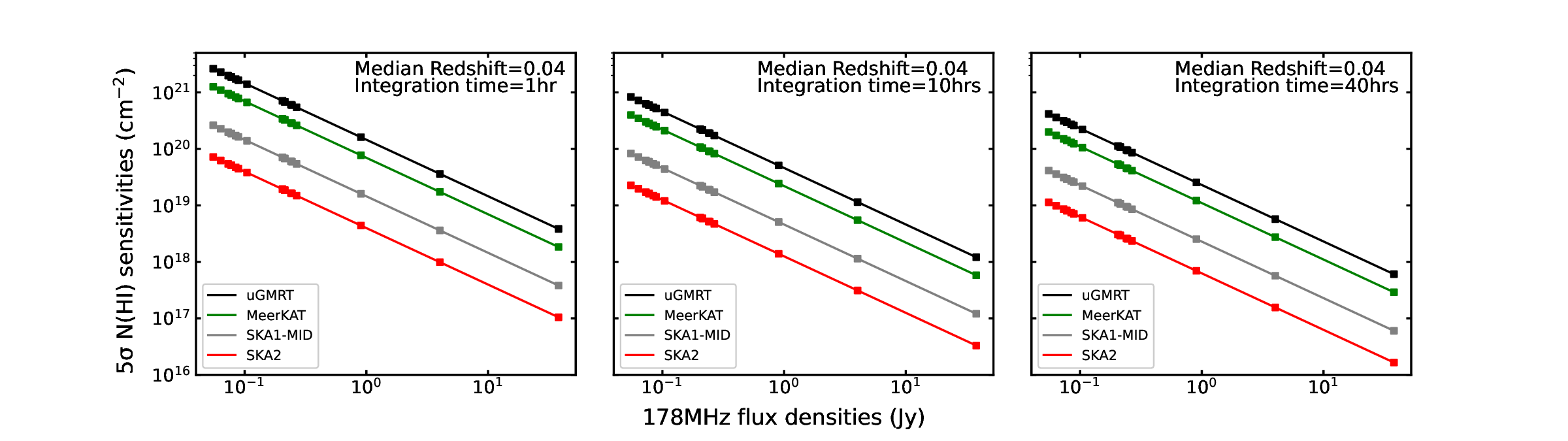}
    \caption{5$\sigma$ N(H~{\sc i}) absorption sensitivities for MOHEGs-type sources for SKA1-MID (grey line), SKA2 (red line) and pathfinder instruments, MeerKAT (green line), and uGMRT (black line) for 1hr (Panel 1), 10hrs (Panel 2) and 40hrs (Panel 3) integration times for a range of flux densities at 178 MHz. 
}
\label{fig:NHIs}
\end{figure*}

Galaxies like radio MOHEGs are expected to have lower H~{\sc i} column densities, which make it difficult to detect them in emission surveys with current radio instruments. H~{\sc i} absorption studies against bright radio continuum background can track down the neutral hydrogen in galaxies, which otherwise could not be seen in H~{\sc i} emission observations due to the constraints of achievable sensitivities at larger distances \citep{Morganti2018,2022Dutta}.    
H~{\sc i} surveys offer substantial advantages to trace star formation and galaxy evolution from the nearby to the distant universe due to their independence from the effects of extinction by dust and color. Hydrogen structures in the inner disk regions, spiral arms, tails, filaments, emission, and absorption profiles tracing the interaction and mergers in galaxies have been regularly detected with the SKA pathfinders instruments such as the ATCA, MeerKAT, uGMRT, VLA, ASKAP, WSRT. \citep{Morganti2018,2020Koribalski}. H~{\sc i} absorption surveys for MOHEG-type sources will enlarge our sample and help better understand the hydrogen gas picture in such galaxies \citep{2023Wagh}.

\begin{figure*}
 \centering
    \includegraphics[width=2.2\columnwidth]{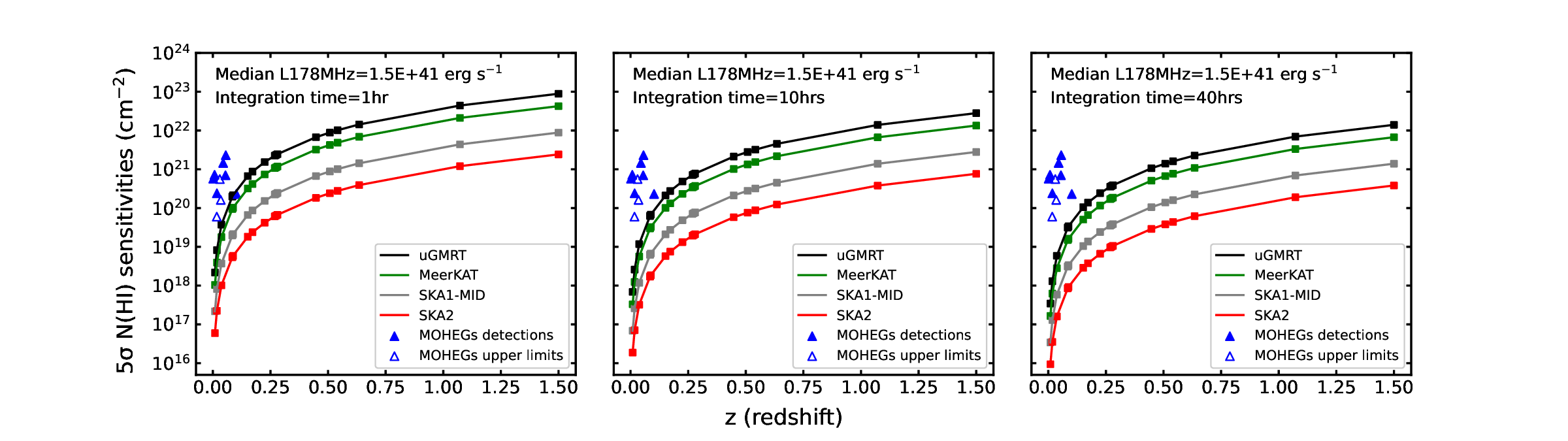}
\caption{5$\sigma$ N(H~{\sc i}) absorption sensitivities for MOHEGs-type sources for SKA1-MID (grey line), SKA2 (red line) and pathfinder instruments, MeerKAT (green line), and uGMRT (black line) for 1hr (Panel 1), 10hrs (Panel 2) and 40hrs (Panel 3) integration times for a range of redshifts. The radio MOHEGs sample is plotted in blue triangles.
}
\label{fig:NHIz}
\end{figure*}

\par The sensitivity for H~{\sc i} detections in MOHEGs-type galaxies at different flux densities, observation time, and redshifts are plotted in Figures \ref{fig:NHIs} and \ref{fig:NHIz} for SKA1-MID, SKA2, MeerKAT and uGMRT. H~{\sc i} absorption limits depend on the strength of the background continuum F$_{1.4}$ GHz flux density, the SEFD (System Equivalent Flux Density) of the radio telescopes, total on-source integration time, and the FWHM, halfwidth of the spectral line. 
 Figure \ref{fig:NHIs} shows N(H~{\sc i}) absorption detection limits based on extrapolated flux densities at a median redshift for different instruments at different integration times. Using the 178 MHz fluxes of the radio MOHEGs \citep{Ogle2010}, and we consider that the MOHEG-type sources would have the same luminosities at five times the luminosity distance as of the current sources, we derived the respective 178 MHz fluxes. We then extrapolated these flux densities at a redshifted H~{\sc i} observing line frequency at the median redshift of the total sample using a power law scaling of -0.7 spectral index (typical value for radio galaxies). Further, taking these flux densities as the background continuum strength values, we calculate 5$\sigma$ sensitivity limits on H~{\sc i} detections for different instruments at different integration times using Equation \ref{eq:4}.  The theoretical RMS noise was calculated for a given integration time at 5 km/s channel width for all the instruments \citep{2015Blyth,2021MAaddox,2019Braun}. With a strong background continuum and reasonable integration times up to $\sim$10 hrs, SKA1-MID and SKA2 bands will detect sources with N(H~{\sc i}) column densities as low as $\sim$ 10$^{17}$ cm$^{-2}$, as compared to MeerKAT and uGMRT. 
The column density N(H~{\sc i}) absorption detection limits as a function of redshift and integration times are shown in Figure \ref{fig:NHIz} along with the radio MOHEGs sample plotted in blue triangles. To generate these plots, the median luminosity at 178 MHz was extrapolated to the corresponding redshifted H~{\sc i} line frequencies. These plots suggest that in shallow surveys with a reasonable integration time of $\sim$1 hrs, SKA1-MID and SKA2 will efficiently detect N(H~{\sc i}) column densities as low as $\sim$ 10$^{21}$ cm$^{-2}$ upto redshift of z $\sim$ 1.5 for a bright population of background sources, thereby probing the cosmic noon regime where star formation activities show a peak. Whereas, focused ($\sim$10) and deep ($\sim$40 hrs) pointed observations will detect as low as $\sim$ 10$^{20}$ cm$^{-2}$ column densities at z $\sim$ 1.5 and  $\sim$ 10$^{17}$ cm$^{-2}$ column densities at z $\sim$ 0 range.
Thus, a new class of low-column-density features can be detected in galaxies with different morphologies residing in diverse environments. This will provide unprecedented information on the gas interaction in galaxies, their evolution, and the impact of the surrounding environment on their capacity for star formation. SKA will efficiently detect low column density features tracing the complete gas cycle from disks to the IGM along with tails and filaments indicating the phase (pre-, post-, ongoing) of merger events and AGN-induced feedback mechanisms. More importantly, it will also provide the complete data to trace the full picture of the H~{\sc i} universe from the current epoch up to the cosmic noon in a reasonable observation time and allow us to probe the universe's star formation history.
\vspace{0.05cm}

\section{Summary and Conclusion} \label{section5}
In this work, we presented a study of hydrogen gas in a sample of radio MOHEGs, focusing on studying the distribution and kinematics of H~{\sc i}  and its influence on star formation relations using H~{\sc i} 21-cm absorption line in this sample and prospects with the SKA and pathfinders for conducting such studies. The main results of this work are summarized below.
\begin{itemize}
    \item  The radio MOHEGs exhibit a deficiency in atomic hydrogen with a median value of less than equal to  5.65 x 10$^{20}$ cm$^{-2}$ as compared to merging galaxies at similar redshifts. However, this can be an order of magnitude higher if the atomic gas predominantly coexists at the same temperature with the unusually warmer molecular gas.
  \item Molecular fraction in the radio MOHEGs is dominated by the H$_2$ gas, with median  $\textit{f}_{\mathrm{{H_2}}}$ $\leq$ 0.96.
    \item There exists no clear correlation between H~{\sc i} gas or total hydrogen (H~{\sc i} + H$_2$) gas column densities and the physical properties of the galaxy such as nearest neighbor distances and radio jet luminosities. The total hydrogen gas densities in the radio MOHEGs provide a complete estimate of gas reservoirs available for star formation, and a possible weak correlation with the SFRs is anticipated in our sample.
 

    \item For our sample, SFRs show a possible, albeit weak, positive correlation with nearest neighbor distances, implying that interactions can influence star formation activities. 
    \item Radio MOHEGs show indications of lower surface densities of star formation compared to star-forming galaxies and ETGs. Also, they show an indication of an offset from the power-law trends established for star-forming galaxies and ETGs. 
    \item We can detect lower H~{\sc i} column densities (10$^{17}$ cm$^{-2}$) through H~{\sc i} absorption surveys of SKA1-MID and SKA2 bands. It will generate statistical data to improve constraints on the evolution of SFRs at cosmic time scales and probe the universe's star formation history from the current epoch up to z $\sim$ 2 and beyond for MOHEG-type sources.
    \end{itemize}
    \par Radio MOHEGs belong to cluster environments or interacting groups/pairs; feedback from neighboring galaxy collisions and tidal interactions may impact the ISM. It could be the reason for excess H$_2$ gas. Also, feedback from such interactions can heat the H$_2$ gas. The density of environments where these galaxies reside also plays an important role in their interaction with the ISM. Quantifying the interaction effects and their impact on the H$_2$ excitation will give us more insights into the feedback through galaxy interactions in these galaxies. Radio jet luminosities are not a good measure of energy transferred into the ISM via jets because there is significant Doppler enhancement in relativistic jets. The study of the isotropic properties of the material thrown out from the jets to radio lobes gives us a better estimate of the kinetic luminosities of the jet. Deep X-ray observations and radio observations of the lobe structures can emphasize better estimations of the plasma states and, hence, the jet kinetic luminosities \citep{2005Punsly}. Calculating kinetic luminosities of the radio MOHEGs will be consequential in quantifying the impact of AGN-jet feedback on the molecular and atomic gas phases. 
    The cold  H$_2$ ($<$100 K) gas distribution is not systematically studied for radio MOHEGs sample; the molecular fraction derived from CO observations will help better understand the role of cold H$_2$ in the star formation processes. Understanding the neutral and molecular hydrogen gas is crucial in understanding galaxy evolution and their star formation activities. Reinforcing the weak correlations with a bigger sample for radio MOHEGs would help us better understand the kinematics of such sources. Radio MOHEGs are interesting objects for examining the interplay between SFRs and feedback processes within radio galaxies.  
\begin{acknowledgments}
We thank the anonymous reviewer for providing valuable comments that enhanced the quality of this paper. SW, NR, MP, AM, CM, BG gratefully acknowledge the support of the Indo-French Centre for the Promotion of Advanced Research (Centre Franco-Indien pour la Promotion de la Recherche Avancée) under project 6504-3.
\end{acknowledgments}

%

\vspace{5mm}





\appendix

\section{Details of H~{\sc i} sensitivity calculations}

Using the following equation, we calculate each instrument's N(H~{\sc i}) absorption sensitivities using the following equation.

  \begin{equation} \label{eq:4}
    \mathrm{N_{HI}} = 1.82 \times 10^{18} \cdot \frac{\mathrm{T_s}}{f\mathrm{_c}} \cdot   \mathrm{\frac{5(RMS)}{S_{1.4}}} \cdot \sqrt{2 \pi}  \sigma_{l} \cdot \sqrt{\frac{\Delta v\mathrm{_{ch}}}{\sigma_{l} }} \mathrm{cm^{-2}}
\end{equation}
T$\mathrm{_s}$ is the spin temperature of H~{\sc i} gas assumed to be 100 K \citep{2019Dutta}.
 \newline $f\mathrm{_c}$ is the fraction of the background radio source covered by the absorbing gas taken as 1.
  \newline RMS is the noise per 5 km/s channel.
 \newline $S_{1.4}$ is the flux density value at 1.4 GHz
 \newline $\sigma_l$ = FWHM / 2.35, this is the H~{\sc i} linewidth; FWHM is assumed to be 100 km s$^{-1}$; because the maximum FWHM for MOHEGs sources is around 100 km s$^{-1}$.
 \newline The extra factor of $\sqrt{\frac{\Delta v\mathrm{_{ch}}}{\sigma l}}$ comes into the picture as we are calculating the sensitivities after smoothing the spectra to a spectral resolution of H~{\sc i} linewidths. $\Delta$ v$\mathrm{_{ch}}$ is the channel width, i.e., 5 km s$^{-1}$
 \begin{table*}
  
 \caption{H~{\sc i} absorption sensitivities for different telescopes}
    \centering
    \begin{tabular}{|c|c|c|c|c|}
        \hline
         Instrument & RMS  & Total integration time & N$\mathrm{_{HI}}$$_{(5\sigma})$  &Reference for RMS\\
          & mJy/beam/5km/s & hours & cm$^{-2}$&\\
          \hline
          uGMRT & 1.044 & 1 & 8.77E+22& *\\
          MeerKAT & 0.500 & 1 & 4.20E+22&\citep{2021MAaddox}\\
          SKA1-MID & 0.104 & 1 & 8.75E+21&\citep{2019Braun}\\
          SKA2 & 0.009 & 1&  2.43E+21&\citep{2015Blyth}\\
          \hline
     \end{tabular}
    
        \small     Notes-  5$\sigma$ sensitivities calculated using minimum background flux density of 0.39~mJy-beam$^{-1}$.  *\url{http://www.ncra.tifr.res.in:8081/~secr-ops/etc/rms/rms.html}

     \label{table4}
 \end{table*}

\bibliography{sample631}{}

\begin{thebibliography}{}
\expandafter\ifx\csname natexlab\endcsname\relax\def\natexlab#1{#1}\fi
\providecommand{\url}[1]{\href{#1}{#1}}
\providecommand{\dodoi}[1]{doi:~\href{http://doi.org/#1}{\nolinkurl{#1}}}
\providecommand{\doeprint}[1]{\href{http://ascl.net/#1}{\nolinkurl{http://ascl.net/#1}}}
\providecommand{\doarXiv}[1]{\href{https://arxiv.org/abs/#1}{\nolinkurl{https://arxiv.org/abs/#1}}}

\bibitem[{{Alladin}(1975)}]{1975Alladin}
{Alladin}, S.~M. 1975, Bulletin of the Astronomical Society of India, 3, 5

\bibitem[{{Allison} {et~al.}(2014){Allison}, {Sadler}, \& {Meekin}}]{2014Allison}
{Allison}, J.~R., {Sadler}, E.~M., \& {Meekin}, A.~M. 2014, \mnras, 440, 696, \dodoi{10.1093/mnras/stu289}

\bibitem[{{Antonuccio-Delogu} \& {Silk}(2008)}]{2008Antonuccio}
{Antonuccio-Delogu}, V., \& {Silk}, J. 2008, \mnras, 389, 1750, \dodoi{10.1111/j.1365-2966.2008.13663.x}

\bibitem[{{Baars} {et~al.}(1977){Baars}, {Genzel}, \& et~al.}]{Baars1977}
{Baars}, J.~M., {Genzel}, R., \& et~al. 1977, A\&A, 61, 99, \dodoi{1977A&A....61...99B}

\bibitem[{{Bergeron} \& {Boiss{\'e}}(1991)}]{1991Bergeon}
{Bergeron}, J., \& {Boiss{\'e}}, P. 1991, \aap, 243, 344

\bibitem[{{Blitz} \& {Rosolowsky}(2004)}]{2004Blitz}
{Blitz}, L., \& {Rosolowsky}, E. 2004, \apjl, 612, L29, \dodoi{10.1086/424661}

\bibitem[{{Blitz} \& {Rosolowsky}(2006)}]{2006Blitz}
---. 2006, \apj, 650, 933, \dodoi{10.1086/505417}

\bibitem[{{Blyth} {et~al.}(2015){Blyth}, {van der Hulst}, {Verheijen}, {Catinella}, {Fraternali}, {Haynes}, {Hess}, {Koribalski}, {Lagos}, {Meyer}, {Obreschkow}, {Popping}, {Power}, {Verdes-Montenegro}, \& {Zwaan}}]{2015Blyth}
{Blyth}, S., {van der Hulst}, T.~M., {Verheijen}, M.~A.~W., {et~al.} 2015, in Advancing Astrophysics with the Square Kilometre Array (AASKA14), 128, \dodoi{10.22323/1.215.0128}

\bibitem[{{Boselli} {et~al.}(2022){Boselli}, {Fossati}, \& {Sun}}]{2022Boselli}
{Boselli}, A., {Fossati}, M., \& {Sun}, M. 2022, \aapr, 30, 3, \dodoi{10.1007/s00159-022-00140-3}

\bibitem[{{Braun} {et~al.}(2019){Braun}, {Bonaldi}, {Bourke}, {Keane}, \& {Wagg}}]{2019Braun}
{Braun}, R., {Bonaldi}, A., {Bourke}, T., {Keane}, E., \& {Wagg}, J. 2019, arXiv e-prints, arXiv:1912.12699, \dodoi{10.48550/arXiv.1912.12699}

\bibitem[{{Brown} {et~al.}(2017){Brown}, {Catinella}, {Cortese}, {Lagos}, {Dav{\'e}}, {Kilborn}, {Haynes}, {Giovanelli}, \& {Rafieferantsoa}}]{2017Brown}
{Brown}, T., {Catinella}, B., {Cortese}, L., {et~al.} 2017, \mnras, 466, 1275, \dodoi{10.1093/mnras/stw2991}

\bibitem[{{Calzetti} {et~al.}(2010){Calzetti}, {Wu}, {Hong}, {Kennicutt}, {Lee}, {Dale}, {Engelbracht}, {van Zee}, {Draine}, {Hao}, {Gordon}, {Moustakas}, {Murphy}, {Regan}, {Begum}, {Block}, {Dalcanton}, {Funes}, {Gil de Paz}, {Johnson}, {Sakai}, {Skillman}, {Walter}, {Weisz}, {Williams}, \& {Wu}}]{2010Calzetti}
{Calzetti}, D., {Wu}, S.~Y., {Hong}, S., {et~al.} 2010, \apj, 714, 1256, \dodoi{10.1088/0004-637X/714/2/1256}

\bibitem[{{Chandola} {et~al.}(2013){Chandola}, {Gupta}, \& {Saikia}}]{2013MNRAS.429.2380C}
{Chandola}, Y., {Gupta}, N., \& {Saikia}, D.~J. 2013, \mnras, 429, 2380, \dodoi{10.1093/mnras/sts499}

\bibitem[{{Chung} {et~al.}(2009){Chung}, {van Gorkom}, {Kenney}, {Crowl}, \& {Vollmer}}]{2009Chung}
{Chung}, A., {van Gorkom}, J.~H., {Kenney}, J. D.~P., {Crowl}, H., \& {Vollmer}, B. 2009, \aj, 138, 1741, \dodoi{10.1088/0004-6256/138/6/1741}

\bibitem[{{Cielo} {et~al.}(2018){Cielo}, {Babul}, {Antonuccio-Delogu}, {Silk}, \& {Volonteri}}]{2018Cielo}
{Cielo}, S., {Babul}, A., {Antonuccio-Delogu}, V., {Silk}, J., \& {Volonteri}, M. 2018, A\&A, 617, A58, \dodoi{10.48550/arXiv.1801.04276}

\bibitem[{{Clark}(1965)}]{1965Clark}
{Clark}, B.~G. 1965, \apj, 142, 1398, \dodoi{10.1086/148426}

\bibitem[{{Cluver} {et~al.}(2013){Cluver}, {Appleton}, {Ogle}, {Jarrett}, {Rasmussen}, {Lisenfeld}, {Guillard}, {Verdes-Montenegro}, {Antonucci}, {Bitsakis}, {Charmandaris}, {Boulanger}, {Egami}, {Xu}, \& {Yun}}]{2013Cluver}
{Cluver}, M.~E., {Appleton}, P.~N., {Ogle}, P., {et~al.} 2013, \apj, 765, 93, \dodoi{10.1088/0004-637X/765/2/93}

\bibitem[{{Colgan} {et~al.}(1990){Colgan}, {Salpeter}, \& {Terzian}}]{1990ApJ...351..503C}
{Colgan}, S. W.~J., {Salpeter}, E.~E., \& {Terzian}, Y. 1990, \apj, 351, 503, \dodoi{10.1086/168488}

\bibitem[{{Cox} \& {Smith}(1974)}]{1974Cox}
{Cox}, D.~P., \& {Smith}, B.~W. 1974, \apjl, 189, L105, \dodoi{10.1086/181476}

\bibitem[{{Curran} {et~al.}(2016){Curran}, {Duchesne}, {Divoli}, \& {Allison}}]{2016Curran}
{Curran}, S.~J., {Duchesne}, S.~W., {Divoli}, A., \& {Allison}, J.~R. 2016, \mnras, 462, 4197, \dodoi{10.1093/mnras/stw1938}

\bibitem[{{Davis} {et~al.}(2014){Davis}, {Young}, {Crocker}, {Bureau}, {Blitz}, {Alatalo}, {Emsellem}, {Naab}, {Bayet}, {Bois}, {Bournaud}, {Cappellari}, {Davies}, {de Zeeuw}, {Duc}, {Khochfar}, {Krajnovi{\'c}}, {Kuntschner}, {McDermid}, {Morganti}, {Oosterloo}, {Sarzi}, {Scott}, {Serra}, \& {Weijmans}}]{2014Davis}
{Davis}, T.~A., {Young}, L.~M., {Crocker}, A.~F., {et~al.} 2014, \mnras, 444, 3427, \dodoi{10.1093/mnras/stu570}

\bibitem[{{D{\'e}nes} {et~al.}(2016){D{\'e}nes}, {Kilborn}, {Koribalski}, \& {Wong}}]{2016Denes}
{D{\'e}nes}, H., {Kilborn}, V.~A., {Koribalski}, B.~S., \& {Wong}, O.~I. 2016, \mnras, 455, 1294, \dodoi{10.1093/mnras/stv2391}

\bibitem[{{Dutta}(2019)}]{2019Dutta}
{Dutta}, R. 2019, Journal of Astrophysics and Astronomy, 40, 41, \dodoi{10.1007/s12036-019-9610-5}

\bibitem[{{Dutta} {et~al.}(2019){Dutta}, {Srianand}, \& {Gupta}}]{2019MNRASDutta}
{Dutta}, R., {Srianand}, R., \& {Gupta}, N. 2019, \mnras, 489, 1099, \dodoi{10.1093/mnras/stz2178}

\bibitem[{{Dutta} {et~al.}(2022){Dutta}, {Kurapati}, {Aditya}, {Bait}, {Das}, {Dutta}, {Indulekha}, {Nandakumar}, {Patra}, {Roy}, \& {Roychowdhury}}]{2022Dutta}
{Dutta}, R., {Kurapati}, S., {Aditya}, J.~N.~H.~S., {et~al.} 2022, Journal of Astrophysics and Astronomy, 43, 103, \dodoi{10.1007/s12036-022-09875-y}

\bibitem[{{Dwarakanath} {et~al.}(1995){Dwarakanath}, {Owen}, \& {van Gorkom}}]{1995ApJ...442L...1D}
{Dwarakanath}, K.~S., {Owen}, F.~N., \& {van Gorkom}, J.~H. 1995, \apjl, 442, L1, \dodoi{10.1086/187801}

\bibitem[{{Emonts} {et~al.}(2010){Emonts}, {Morganti}, {Struve}, {Oosterloo}, {van Moorsel}, {Tadhunter}, {van der Hulst}, {Brogt}, {Holt}, \& {Mirabal}}]{2010Emonts}
{Emonts}, B.~H.~C., {Morganti}, R., {Struve}, C., {et~al.} 2010, \mnras, 406, 987, \dodoi{10.1111/j.1365-2966.2010.16706.x}

\bibitem[{{Fabian}(1994)}]{Fabian1994}
{Fabian}, A.~C. 1994, ARA\&A, 32, \dodoi{https://doi.org/10.1146/annurev.aa.32.090194.001425}

\bibitem[{{Fanaroff} \& {Riley}(1974)}]{1974Fanaroff}
{Fanaroff}, B.~L., \& {Riley}, J.~M. 1974, \mnras, 167, 31P, \dodoi{10.1093/mnras/167.1.31P}

\bibitem[{{Federman} {et~al.}(1979){Federman}, {Glassgold}, \& {Kwan}}]{1979Federman}
{Federman}, S.~R., {Glassgold}, A.~E., \& {Kwan}, J. 1979, \apj, 227, 466, \dodoi{10.1086/156753}

\bibitem[{{Foley} {et~al.}(2018){Foley}, {Cazaux}, {Egorov}, {Boschman}, {Hoekstra}, \& {Schlath{\"o}lter}}]{2018Foley}
{Foley}, N., {Cazaux}, S., {Egorov}, D., {et~al.} 2018, \mnras, 479, 649, \dodoi{10.1093/mnras/sty1528}

\bibitem[{{Fujita} {et~al.}(2022){Fujita}, {Kawakatu}, \& {Nagai}}]{Fujita2022}
{Fujita}, Y., {Kawakatu}, N., \& {Nagai}, H. 2022, \apj, 924, 24, \dodoi{10.3847/1538-4357/ac31a6}

\bibitem[{{Gentile} {et~al.}(2007){Gentile}, {Rodr{\'\i}guez}, {Taylor}, {Giovannini}, {Allen}, {Lane}, \& {Kassim}}]{2007Gentile}
{Gentile}, G., {Rodr{\'\i}guez}, C., {Taylor}, G.~B., {et~al.} 2007, \apj, 659, 225, \dodoi{10.1086/512479}

\bibitem[{{Ger{\'e}b} {et~al.}(2015){Ger{\'e}b}, {Maccagni}, {Morganti}, \& {Oosterloo}}]{2015GER}
{Ger{\'e}b}, K., {Maccagni}, F.~M., {Morganti}, R., \& {Oosterloo}, T.~A. 2015, \aap, 575, A44, \dodoi{10.1051/0004-6361/201424655}

\bibitem[{{Guglielmo} {et~al.}(2015){Guglielmo}, {Poggianti}, {Moretti}, {Fritz}, {Calvi}, {Vulcani}, {Fasano}, \& {Paccagnella}}]{SFR2015}
{Guglielmo}, V., {Poggianti}, B.~M., {Moretti}, A., {et~al.} 2015, \mnras, 450, 2749, \dodoi{10.1093/mnras/stv757}

\bibitem[{{Guillard} {et~al.}(2012){Guillard}, {Ogle}, {Emonts}, {Appleton}, {Morganti}, {Tadhunter}, {Oosterloo}, {Evans}, \& {Evans}}]{2012ApJ...747...95G}
{Guillard}, P., {Ogle}, P.~M., {Emonts}, B.~H.~C., {et~al.} 2012, \apj, 747, 95, \dodoi{10.1088/0004-637X/747/2/95}

\bibitem[{{Hollenbach} {et~al.}(1971){Hollenbach}, {Werner}, \& {Salpeter}}]{1971Hollenbach}
{Hollenbach}, D.~J., {Werner}, M.~W., \& {Salpeter}, E.~E. 1971, \apj, 163, 165, \dodoi{10.1086/150755}

\bibitem[{{Jaffe} \& {McNamara}(1994)}]{1994Jaffe}
{Jaffe}, W., \& {McNamara}, B.~R. 1994, \apj, 434, 110, \dodoi{10.1086/174708}

\bibitem[{{Kanekar} {et~al.}(2011){Kanekar}, {Braun}, \& {Roy}}]{2011Kanekar}
{Kanekar}, N., {Braun}, R., \& {Roy}, N. 2011, \apjl, 737, L33, \dodoi{10.1088/2041-8205/737/2/L33}

\bibitem[{{Kawata} \& {Mulchaey}(2008)}]{2008Kawata}
{Kawata}, D., \& {Mulchaey}, J.~S. 2008, \apjl, 672, L103, \dodoi{10.1086/526544}

\bibitem[{{Kennicutt}(1998{\natexlab{a}})}]{Ken1998}
{Kennicutt}, Robert~C., J. 1998{\natexlab{a}}, \apj, 498, 541, \dodoi{10.1086/305588}

\bibitem[{{Kennicutt}(1998{\natexlab{b}})}]{1998Kennicutt}
---. 1998{\natexlab{b}}, \araa, 36, 189, \dodoi{10.1146/annurev.astro.36.1.189}

\bibitem[{{Kennicutt} \& {De Los Reyes}(2021)}]{2021Kennicutt}
{Kennicutt}, Robert~C., J., \& {De Los Reyes}, M. A.~C. 2021, \apj, 908, 61, \dodoi{10.3847/1538-4357/abd3a2}

\bibitem[{{Knapp} {et~al.}(1978){Knapp}, {Kerr}, \& {Williams}}]{1978Knapp}
{Knapp}, G.~R., {Kerr}, F.~J., \& {Williams}, B.~A. 1978, \apj, 222, 800, \dodoi{10.1086/156199}

\bibitem[{{Koribalski} {et~al.}(2020){Koribalski}, {Staveley-Smith}, {Westmeier}, {Serra}, {Spekkens}, {Wong}, {Lee-Waddell}, {Lagos}, {Obreschkow}, {Ryan-Weber}, {Zwaan}, {Kilborn}, {Bekiaris}, {Bekki}, {Bigiel}, {Boselli}, {Bosma}, {Catinella}, {Chauhan}, {Cluver}, {Colless}, {Courtois}, {Crain}, {de Blok}, {D{\'e}nes}, {Duffy}, {Elagali}, {Fluke}, {For}, {Heald}, {Henning}, {Hess}, {Holwerda}, {Howlett}, {Jarrett}, {Jones}, {Jones}, {J{\'o}zsa}, {Jurek}, {J{\"u}tte}, {Kamphuis}, {Karachentsev}, {Kerp}, {Kleiner}, {Kraan-Korteweg}, {L{\'o}pez-S{\'a}nchez}, {Madrid}, {Meyer}, {Mould}, {Murugeshan}, {Norris}, {Oh}, {Oosterloo}, {Popping}, {Putman}, {Reynolds}, {Rhee}, {Robotham}, {Ryder}, {Schr{\"o}der}, {Shao}, {Stevens}, {Taylor}, {van{\^A} der Hulst}, {Verdes-Montenegro}, {Wakker}, {Wang}, {Whiting}, {Winkel}, \& {Wolf}}]{2020Koribalski}
{Koribalski}, B.~S., {Staveley-Smith}, L., {Westmeier}, T., {et~al.} 2020, \apss, 365, 118, \dodoi{10.1007/s10509-020-03831-4}

\bibitem[{{Lada}(2005)}]{SFR2005}
{Lada}, C.~J. 2005, Progress of Theoretical Physics Supplement, 158, 1, \dodoi{10.1143/PTPS.158.1}

\bibitem[{{Lanz} {et~al.}(2016){Lanz}, {Ogle}, {Alatalo}, \& {Appleton}}]{2016Lanz}
{Lanz}, L., {Ogle}, P.~M., {Alatalo}, K., \& {Appleton}, P.~N. 2016, \apj, 826, 29, \dodoi{10.3847/0004-637X/826/1/29}

\bibitem[{{Le Bourlot} {et~al.}(1999){Le Bourlot}, {Pineau des For{\^e}ts}, \& {Flower}}]{1999Le}
{Le Bourlot}, J., {Pineau des For{\^e}ts}, G., \& {Flower}, D.~R. 1999, \mnras, 305, 802, \dodoi{10.1046/j.1365-8711.1999.02497.x}

\bibitem[{{Leroy} {et~al.}(2008){Leroy}, {Walter}, {Brinks}, {Bigiel}, {de Blok}, {Madore}, \& {Thornley}}]{2008Leroy}
{Leroy}, A.~K., {Walter}, F., {Brinks}, E., {et~al.} 2008, \aj, 136, 2782, \dodoi{10.1088/0004-6256/136/6/2782}

\bibitem[{{Li} {et~al.}(2015){Li}, {Ostriker}, {Cen}, {Bryan}, \& {Naab}}]{201Li}
{Li}, M., {Ostriker}, J.~P., {Cen}, R., {Bryan}, G.~L., \& {Naab}, T. 2015, \apj, 814, 4, \dodoi{10.1088/0004-637X/814/1/4}

\bibitem[{{Maddox} {et~al.}(2021){Maddox}, {Frank}, {Ponomareva}, {Jarvis}, {Adams}, {Dav{\'e}}, {Oosterloo}, {Santos}, {Blyth}, {Glowacki}, {Kraan-Korteweg}, {Mulaudzi}, {Namumba}, {Prandoni}, {Rajohnson}, {Spekkens}, {Adams}, {Bowler}, {Collier}, {Heywood}, {Sekhar}, \& {Taylor}}]{2021MAaddox}
{Maddox}, N., {Frank}, B.~S., {Ponomareva}, A.~A., {et~al.} 2021, \aap, 646, A35, \dodoi{10.1051/0004-6361/202039655}

\bibitem[{{Mahony} {et~al.}(2013){Mahony}, {Morganti}, {Emonts}, {Oosterloo}, \& {Tadhunter}}]{2013Mahony}
{Mahony}, E.~K., {Morganti}, R., {Emonts}, B.~H.~C., {Oosterloo}, T.~A., \& {Tadhunter}, C. 2013, \mnras, 435, L58, \dodoi{10.1093/mnrasl/slt094}

\bibitem[{{Mahony} {et~al.}(2016){Mahony}, {Oonk}, {Morganti}, {Tadhunter}, {Bessiere}, {Short}, {Emonts}, \& {Oosterloo}}]{2016Mahony}
{Mahony}, E.~K., {Oonk}, J.~B.~R., {Morganti}, R., {et~al.} 2016, \mnras, 455, 2453, \dodoi{10.1093/mnras/stv2456}

\bibitem[{{Maiolino}(2017)}]{Maiolino2017}
{Maiolino}, R. 2017, ApJ, 544, 202, \dodoi{https://doi.org/10.48550/arXiv.1703.08587}

\bibitem[{{Man} \& {Sirio}(2018)}]{Man2018}
{Man}, A.~M., \& {Sirio}, B. 2018, Nature Astronomy, 2, 695, \dodoi{https://doi.org/10.1038/s41550-018-0558-1}

\bibitem[{{Maret} {et~al.}(2009){Maret}, {Bergin}, {Neufeld}, {Green}, {Watson}, {Harwit}, {Kristensen}, {Melnick}, {Sonnentrucker}, {Tolls}, {Werner}, {Willacy}, \& {Yuan}}]{2009Maret}
{Maret}, S., {Bergin}, E.~A., {Neufeld}, D.~A., {et~al.} 2009, \apj, 698, 1244, \dodoi{10.1088/0004-637X/698/2/1244}

\bibitem[{{Martizzi} {et~al.}(2015){Martizzi}, {Faucher-Gigu{\`e}re}, \& {Quataert}}]{2015Martizzi}
{Martizzi}, D., {Faucher-Gigu{\`e}re}, C.-A., \& {Quataert}, E. 2015, \mnras, 450, 504, \dodoi{10.1093/mnras/stv562}

\bibitem[{{Mel{\'e}ndez} {et~al.}(2014){Mel{\'e}ndez}, {Mushotzky}, {Shimizu}, {Barger}, \& {Cowie}}]{2014MEL}
{Mel{\'e}ndez}, M., {Mushotzky}, R.~F., {Shimizu}, T.~T., {Barger}, A.~J., \& {Cowie}, L.~L. 2014, \apj, 794, 152, \dodoi{10.1088/0004-637X/794/2/152}

\bibitem[{{Morganti} {et~al.}(2003){Morganti}, {Oosterloo}, {Emonts}, {van der Hulst}, \& {Tadhunter}}]{Morganti2003}
{Morganti}, R., {Oosterloo}, T.~A., {Emonts}, B.~H.~C., {van der Hulst}, J.~M., \& {Tadhunter}, C.~N. 2003, \apjl, 593, L69, \dodoi{10.1086/378219}

\bibitem[{{Morganti} \& {Osterloo}(2018)}]{Morganti2018}
{Morganti}, R., \& {Osterloo}, T. 2018, A\&A, \dodoi{https://doi.org/10.48550/arXiv.1807.01475}

\bibitem[{{Morganti} {et~al.}(2005){Morganti}, {Tadhunter}, \& {Oosterloo}}]{Morganti2005}
{Morganti}, R., {Tadhunter}, C.~N., \& {Oosterloo}, T.~A. 2005, \aap, 444, L9, \dodoi{10.1051/0004-6361:200500197}

\bibitem[{{Morganti} {et~al.}(2023){Morganti}, {Murthy}, {Oosterloo}, {Blanchard}, {Cook}, {Paragi}, {Orienti}, {Nagai}, \& {Schulz}}]{2023Morganti}
{Morganti}, R., {Murthy}, S., {Oosterloo}, T., {et~al.} 2023, \aap, 678, A42, \dodoi{10.1051/0004-6361/202347117}

\bibitem[{{Moss} {et~al.}(2017){Moss}, {Allison}, {Sadler}, {Urquhart}, {Soria}, {Callingham}, {Curran}, {Musaeva}, {Mahony}, {Glowacki}, {Farrell}, {Bannister}, {Chippendale}, {Edwards}, {Harvey-Smith}, {Heywood}, {Hotan}, {Indermuehle}, {Lenc}, {Marvil}, {McConnell}, {Reynolds}, {Voronkov}, {Wark}, \& {Whiting}}]{2017Moss}
{Moss}, V.~A., {Allison}, J.~R., {Sadler}, E.~M., {et~al.} 2017, \mnras, 471, 2952, \dodoi{10.1093/mnras/stx1679}

\bibitem[{{Naab} \& {Ostriker}(2017)}]{2017Naab}
{Naab}, T., \& {Ostriker}, J.~P. 2017, \araa, 55, 59, \dodoi{10.1146/annurev-astro-081913-040019}

\bibitem[{{Narayanan} {et~al.}(2008){Narayanan}, {Cox}, {Kelly}, {Dav{\'e}}, {Hernquist}, {Di Matteo}, {Hopkins}, {Kulesa}, {Robertson}, \& {Walker}}]{2008Narayanan}
{Narayanan}, D., {Cox}, T.~J., {Kelly}, B., {et~al.} 2008, \apjs, 176, 331, \dodoi{10.1086/533500}

\bibitem[{{Nesvadba} {et~al.}(2010){Nesvadba}, {Boulanger}, {Salom{\'e}}, {Guillard}, {Lehnert}, {Ogle}, {Appleton}, {Falgarone}, \& {Pineau Des Forets}}]{2010Nes}
{Nesvadba}, N.~P.~H., {Boulanger}, F., {Salom{\'e}}, P., {et~al.} 2010, \aap, 521, A65, \dodoi{10.1051/0004-6361/200913333}

\bibitem[{{Noordermeer} {et~al.}(2005){Noordermeer}, {van der Hulst}, {Sancisi}, {Swaters}, \& {van Albada}}]{2005Noordermeer}
{Noordermeer}, E., {van der Hulst}, J.~M., {Sancisi}, R., {Swaters}, R.~A., \& {van Albada}, T.~S. 2005, \aap, 442, 137, \dodoi{10.1051/0004-6361:20053172}

\bibitem[{{Ogle} {et~al.}(2007){Ogle}, {Antonucci}, {Appleton}, \& {Whysong}}]{2007ApJ...668..699O}
{Ogle}, P., {Antonucci}, R., {Appleton}, P.~N., \& {Whysong}, D. 2007, \apj, 668, 699, \dodoi{10.1086/521334}

\bibitem[{{Ogle} {et~al.}(2010){Ogle}, {Boulanger}, {Guillard}, {Evans}, {Antonucci}, {Appleton}, {Nesvadba}, \& {Leipski}}]{Ogle2010}
{Ogle}, P., {Boulanger}, F., {Guillard}, P., {et~al.} 2010, \apj, 724, 1193, \dodoi{10.1088/0004-637X/724/2/1193}

\bibitem[{{Pan} {et~al.}(2018){Pan}, {Lin}, {Hsieh}, {Xiao}, {Gao}, {Ellison}, {Scudder}, {Barrera-Ballesteros}, {Yuan}, {Saintonge}, {Wilson}, {Hwang}, {De Looze}, {Gao}, {Ho}, {Brinks}, {Mok}, {Brown}, {Davis}, {Williams}, {Chung}, {Parsons}, {Bureau}, {Sargent}, {Chung}, {Kim}, {Liu}, {Micha{\l}owski}, \& {Tosaki}}]{2018Pan}
{Pan}, H.-A., {Lin}, L., {Hsieh}, B.-C., {et~al.} 2018, \apj, 868, 132, \dodoi{10.3847/1538-4357/aaeb92}

\bibitem[{{Park} {et~al.}(2023){Park}, {Lee}, {Bialy}, {Burkhart}, {Dawson}, {Heiles}, {Li}, {Murray}, {Nguyen}, {Hafner}, {Rybarczyk}, \& {Stanimirovi{\'c}}}]{2023Park}
{Park}, G., {Lee}, M.-Y., {Bialy}, S., {et~al.} 2023, ApJ, arXiv:2307.09513, \dodoi{10.48550/arXiv.2307.09513}

\bibitem[{{Pearson} {et~al.}(2019){Pearson}, {Wang}, {Alpaslan}, {Baldry}, {Bilicki}, {Brown}, {Grootes}, {Holwerda}, {Kitching}, {Kruk}, \& {van der Tak}}]{2019Pearson}
{Pearson}, W.~J., {Wang}, L., {Alpaslan}, M., {et~al.} 2019, \aap, 631, A51, \dodoi{10.1051/0004-6361/201936337}

\bibitem[{{Peng} {et~al.}(2015){Peng}, {Maiolino}, \& {Cochrane}}]{2015Peng}
{Peng}, Y., {Maiolino}, R., \& {Cochrane}, R. 2015, \nat, 521, 192, \dodoi{10.1038/nature14439}

\bibitem[{{Punsly}(2005)}]{2005Punsly}
{Punsly}, B. 2005, \apjl, 623, L9, \dodoi{10.1086/430140}

\bibitem[{{Rauch}(1998)}]{1998Rauch}
{Rauch}, M. 1998, \araa, 36, 267, \dodoi{10.1146/annurev.astro.36.1.267}

\bibitem[{{Rohlfs} \& {Wilson}(2000)}]{2000Rohlfs}
{Rohlfs}, K., \& {Wilson}, T.~L. 2000, {Tools of radio astronomy}

\bibitem[{{Rupke} \& {Veilleux}(2011)}]{2011Rupke}
{Rupke}, D. S.~N., \& {Veilleux}, S. 2011, \apjl, 729, L27, \dodoi{10.1088/2041-8205/729/2/L27}

\bibitem[{{Saury} {et~al.}(2014){Saury}, {Miville-Desch{\^e}nes}, {Hennebelle}, {Audit}, \& {Schmidt}}]{2014Saury}
{Saury}, E., {Miville-Desch{\^e}nes}, M.~A., {Hennebelle}, P., {Audit}, E., \& {Schmidt}, W. 2014, \aap, 567, A16, \dodoi{10.1051/0004-6361/201321113}

\bibitem[{{Schmidt}(1959)}]{KS1959}
{Schmidt}, M. 1959, \apj, 129, 243, \dodoi{10.1086/146614}

\bibitem[{{Schmidt}(1963)}]{1963Schimdt}
---. 1963, \apj, 137, 758, \dodoi{10.1086/147553}

\bibitem[{{Schulz} {et~al.}(2021){Schulz}, {Morganti}, {Nyland}, {Paragi}, {Mahony}, \& {Oosterloo}}]{2021Schulz}
{Schulz}, R., {Morganti}, R., {Nyland}, K., {et~al.} 2021, \aap, 647, A63, \dodoi{10.1051/0004-6361/202037677}

\bibitem[{{Serra} {et~al.}(2012){Serra}, {Oosterloo}, {Morganti}, {Alatalo}, {Blitz}, {Bois}, {Bournaud}, {Bureau}, {Cappellari}, {Crocker}, {Davies}, {Davis}, {de Zeeuw}, {Duc}, {Emsellem}, {Khochfar}, {Krajnovi{\'c}}, {Kuntschner}, {Lablanche}, {McDermid}, {Naab}, {Sarzi}, {Scott}, {Trager}, {Weijmans}, \& {Young}}]{Serra2012}
{Serra}, P., {Oosterloo}, T., {Morganti}, R., {et~al.} 2012, \mnras, 422, 1835, \dodoi{10.1111/j.1365-2966.2012.20219.x}

\bibitem[{{Shi} {et~al.}(2005){Shi}, {Rieke}, {Hines}, {Neugebauer}, {Blaylock}, {Rigby}, {Egami}, {Gordon}, \& {Alonso-Herrero}}]{2005Shi}
{Shi}, Y., {Rieke}, G.~H., {Hines}, D.~C., {et~al.} 2005, \apj, 629, 88, \dodoi{10.1086/431344}

\bibitem[{{Silk}(2013)}]{Silk2013}
{Silk}, J. 2013, ApJ, 772, \dodoi{https://doi.org/10.48550/arXiv.1305.5840}

\bibitem[{{Stecher} \& {Williams}(1967)}]{1967Stecher}
{Stecher}, T.~P., \& {Williams}, D.~A. 1967, \apjl, 149, L29, \dodoi{10.1086/180047}

\bibitem[{{Struve} \& {Conway}(2010)}]{2010Struve}
{Struve}, C., \& {Conway}, J.~E. 2010, \aap, 513, A10, \dodoi{10.1051/0004-6361/200913572}

\bibitem[{{Thi} {et~al.}(2020){Thi}, {Hocuk}, {Kamp}, {Woitke}, {Rab}, {Cazaux}, \& {Caselli}}]{2020A&A...634A..42T}
{Thi}, W.~F., {Hocuk}, S., {Kamp}, I., {et~al.} 2020, \aap, 634, A42, \dodoi{10.1051/0004-6361/201731746}

\bibitem[{{Toomre}(1974)}]{1974Toomre}
{Toomre}, A. 1974, in The Formation and Dynamics of Galaxies, ed. J.~R. {Shakeshaft}, Vol.~58, 347

\bibitem[{{van Gorkom} {et~al.}(1989){van Gorkom}, {Knapp}, {Ekers}, {Ekers}, {Laing}, \& {Polk}}]{1989AJ.....97..708V}
{van Gorkom}, J.~H., {Knapp}, G.~R., {Ekers}, R.~D., {et~al.} 1989, \aj, 97, 708, \dodoi{10.1086/115016}

\bibitem[{{Veilleux} {et~al.}(2017){Veilleux}, {Bolatto}, {Tombesi}, {Mel{\'e}ndez}, {Sturm}, {Gonz{\'a}lez-Alfonso}, {Fischer}, \& {Rupke}}]{2017Veilleux}
{Veilleux}, S., {Bolatto}, A., {Tombesi}, F., {et~al.} 2017, \apj, 843, 18, \dodoi{10.3847/1538-4357/aa767d}

\bibitem[{{Vidali}(2013)}]{2013Vidali}
{Vidali}, G. 2013, Chemical Reviews, 113, 8752, \dodoi{10.1021/cr400156b}

\bibitem[{{Vollmer}(2013)}]{2013Vollmer}
{Vollmer}, B. 2013, in , Vol.~6, 207, \dodoi{10.1007/978-94-007-5609-0_5}

\bibitem[{{Wagh} {et~al.}(2023){Wagh}, {Pandey-Pommier}, {Roy}, {Rashid}, {Marcowith}, {Roy}, {Muthumariappan}, {Sethuram}, \& {Guiderdoni}}]{2023Wagh}
{Wagh}, S., {Pandey-Pommier}, M., {Roy}, N., {et~al.} 2023, in SF2A-2023: Proceedings of the Annual meeting of the French Society of Astronomy and Astrophysics, 71--74

\bibitem[{{Wang} {et~al.}(2016){Wang}, {Koribalski}, {Serra}, {van der Hulst}, {Roychowdhury}, {Kamphuis}, \& {Chengalur}}]{2016Wang}
{Wang}, J., {Koribalski}, B.~S., {Serra}, P., {et~al.} 2016, \mnras, 460, 2143, \dodoi{10.1093/mnras/stw1099}

\bibitem[{{We{\.z}gowiec} {et~al.}(2012){We{\.z}gowiec}, {Bomans}, {Ehle}, {Chy{\.z}y}, {Urbanik}, {Braine}, \& {Soida}}]{2012We}
{We{\.z}gowiec}, M., {Bomans}, D.~J., {Ehle}, M., {et~al.} 2012, \aap, 544, A99, \dodoi{10.1051/0004-6361/201117652}

\bibitem[{{Winkel} {et~al.}(2017){Winkel}, {Wiesemeyer}, {Menten}, {Sato}, {Brunthaler}, {Wyrowski}, {Neufeld}, {Gerin}, \& {Indriolo}}]{2017Winkel}
{Winkel}, B., {Wiesemeyer}, H., {Menten}, K.~M., {et~al.} 2017, \aap, 600, A2, \dodoi{10.1051/0004-6361/201628597}

\bibitem[{{Wolfe} {et~al.}(2005){Wolfe}, {Gawiser}, \& {Prochaska}}]{2005Wolfe}
{Wolfe}, A.~M., {Gawiser}, E., \& {Prochaska}, J.~X. 2005, \araa, 43, 861, \dodoi{10.1146/annurev.astro.42.053102.133950}

\bibitem[{{Wolfire} {et~al.}(1995){Wolfire}, {Hollenbach}, {McKee}, {Tielens}, \& {Bakes}}]{1995Wolfire}
{Wolfire}, M.~G., {Hollenbach}, D., {McKee}, C.~F., {Tielens}, A.~G.~G.~M., \& {Bakes}, E.~L.~O. 1995, \apj, 443, 152, \dodoi{10.1086/175510}

\bibitem[{{Wolfire} {et~al.}(2003){Wolfire}, {McKee}, {Hollenbach}, \& {Tielens}}]{2003Wolfire}
{Wolfire}, M.~G., {McKee}, C.~F., {Hollenbach}, D., \& {Tielens}, A.~G.~G.~M. 2003, \apj, 587, 278, \dodoi{10.1086/368016}

\bibitem[{{Wong} \& {Blitz}(2002)}]{2002Wong}
{Wong}, T., \& {Blitz}, L. 2002, \apj, 569, 157, \dodoi{10.1086/339287}

\bibitem[{{Wright}(2006)}]{2006Wright}
{Wright}, E.~L. 2006, \pasp, 118, 1711, \dodoi{10.1086/510102}

\bibitem[{{Yun} {et~al.}(1994){Yun}, {Ho}, \& {Lo}}]{1994Yun}
{Yun}, M.~S., {Ho}, P.~T.~P., \& {Lo}, K.~Y. 1994, \nat, 372, 530, \dodoi{10.1038/372530a0}

\end{thebibliography}
\bibliographystyle{aasjournal}



\end{document}